\documentclass[11pt]{article}
\usepackage{amsthm,amsfonts,amsmath,amssymb}
\usepackage{graphicx}
\theoremstyle{plain}

\newtheorem{thm}{Theorem}[section]
\newtheorem{lem}[thm]{Lemma}

\theoremstyle{definition}

\newtheorem{rem}[thm]{Remark}
\theoremstyle{remark}

\def\ts{\textstyle}

\def\K{{\cal K}}
\def\F{{\cal F}}

\newcommand{\eps}{\epsilon}
\newcommand{\al}{\alpha}
\newcommand{\rA}{\rho_A}
\newcommand{\rB}{\rho_B}
\newcommand{\rC}{\rho_C}

\newcommand{\PRE}{Phys. Rev. E.}
\newcommand{\PRL}{Phys. Rev. Lett.}

\DeclareMathOperator{\sn}{sn}
\newcommand{\pA}{\phi_A}
\newcommand{\pB}{\phi_B}

\newcommand{\be}{\begin{equation}}
\newcommand{\ee}{\end{equation}}
\newcommand{\bea}{\begin{eqnarray}}
\newcommand{\eea}{\end{eqnarray}}

\renewcommand{\atop}[2]{\genfrac{}{}{0pt}{}{#1}{#2}}
\newcommand \rhobar{r}
\newcommand \zbar{\bar z}
\newcommand{\cthm}[1]{Theorem~\ref{#1}}

\newcommand{\clem}[1]{Lemma~\ref{#1}}

\newcommand{\cfig}[1]{Figure~\ref{#1}}
\newcommand{\crem}[1]{Remark~\ref{#1}}

\newcommand{\E}{{\mathcal E}}
\newcommand{\zetau}{{\underline\zeta}}
\renewcommand{\S}{{\mathcal S}}
\newcommand{\floor}[1]{\lfloor#1\rfloor}
\newcommand{\Prob}{\mathop{\rm Prob}}

\newcommand\bR{{\mathbb R}}

\newcommand\bZ{{\mathbb Z}}
\newenvironment{proofof}[1]{\medskip\noindent
   \textit{Proof of #1:} }{\hfill \qed\par\medskip}

\numberwithin{equation}{section} 
\title{Phase diagram of the ABC model on an interval}
\date{August 28, 2009}
\author{{ A. Ayyer${}^1$, E. A. Carlen${}^{2}$, J. L. Lebowitz${}^{2,3}$,
P. K. Mohanty${}^4$,}\\
{ D. Mukamel${}^5$ and E. R. Speer${}^2$}\\ \\
{\small $^1$ Institut de Physique Th\'eorique, IPhT, CEA Saclay, and URA 2306, 
CNRS,}\\
{\small 91191 Gif-sur-Yvette Cedex, FRANCE}\\
{\small $^2$ Department of Mathematics, Rutgers University,}\\
{\small Piscataway, NJ 08854-8019 USA}\\
{\small $^3$ Department of Physics, Rutgers University,}\\
{\small Piscataway, NJ 08854 USA}\\
{\small $^4$ TCMP Division, Saha Institute of Nuclear Physics,}\\
{\small 1/AF Bidhan Nagar, 700064 INDIA} \\
{\small $^5$ Department of Physics of Complex Systems,}\\
{\small Weizmann Institute of Science, Rehovot 76100, ISRAEL}}

\begin{document}

\maketitle
\begin{abstract}

The three species asymmetric $ABC$ model was initially defined on a ring by
Evans, Kafri, Koduvely, and Mukamel, and the weakly asymmetric version was
later studied by Clincy, Derrida, and Evans.  Here the latter model is
studied on a one-dimensional lattice of $N$ sites with closed (zero flux)
boundaries.  In this geometry the local particle conserving dynamics
satisfies detailed balance with respect to a canonical Gibbs measure with
long range asymmetric pair interactions.  This generalizes results for the
ring case, where detailed balance holds, and in fact the steady state
measure is known only for the case of equal densities of the different
species: in the latter case the stationary states of the system on a ring
and on an interval are the same.  We prove that in the limit $N\to \infty$
the scaled density profiles are given by (pieces of) the periodic
trajectory of a particle moving in a quartic confining potential.  We
further prove uniqueness of the profiles, i.e., the existence of a single
phase, in all regions of the parameter space (of average densities and
temperature) except at low temperature with all densities equal; in this
case a continuum of phases, differing by translation, coexist. The results
for the equal density case apply also to the system on the ring, and there
extend results of Clincy et al.

\end{abstract}

\section{Introduction}\label{sec:intro}

One dimensional systems play an important role in statistical mechanics.
In addition to their intrinsic interest as models of physical systems in
confined quasi-linear geometries, they are in many cases exactly solvable
and in fact are, with few exceptions, the only exactly solvable many body
systems \cite{Leib,Percus,Privman,Schutz}.  These exact solutions provide
insights into both equilibrium and non-equilibrium collective behavior in
higher dimensions.

An interesting connection between equilibrium and non-equilibrium phase
transitions in one dimension is provided by the ABC model introduced by
Evans et al.~\cite{Evans98} (a model with similar behavior was discussed
in \cite{LBR}): a one dimensional system consisting of three
species of particles, labeled $A, B, C$, on a ring containing $N$ lattice
sites (one may equivalently regard it as a model with two species and empty
sites, or as a ``clock'' model with three states).  We will here typically let
$\alpha=A$, $B$, or $C$ denote a particle type, and make the convention
that $\alpha+1$, $\alpha+2$, $\ldots$ denote the particle types which are
successors to $\alpha$ in the cyclic order $ABC$.  The system evolves by
particle conserving nearest neighbor exchanges with asymmetric rates:
$\alpha \gamma \to \gamma \alpha$ (clockwise) with rate
$q_{\alpha,\gamma}$, where $q_{\alpha,\alpha+1} < q_{\alpha+1,\alpha}$; the
total numbers $N_\alpha$ of particles of each species are conserved and
satisfy $\sum_\alpha N_\alpha=N$.  Evans et al.~argued that this system
will, in the limit $N \to \infty$ with $N_\alpha/N\to r_\alpha$ and with
$r_\alpha>0$ for all $\alpha$, segregate into pure $A$, $B$, and $C$
regions, with rotationally invariant distribution of the phase boundaries.
For the case $q_{AB} = q_{BC} = q_{CA} = q <1$ and
$q_{BA} = q_{CB} = q_{AC} = 1$, the only case we will consider here, they
showed further that when $N_A = N_B = N_C = N/3$ the dynamics satisfies
detailed balance with respect to the Gibbs measure of a certain Hamiltonian
having a long range pair interaction, so that the stationary state is an
equilibrium state.

In a later development Clincy et al.~\cite{CDE} considered a weakly
asymmetric version of this model in which $q = e^{-\beta/N}$; the
stationary state for the equal density case $N_\alpha = N/3$ then becomes a
Gibbs measure of the form $\exp \{-\beta \hat E_N\}$, with energy
$\hat E_N$ equal, up to a constant, to the Hamiltonian of \cite{Evans98}
divided by $N$.  The parameter $\beta$ thus plays the role of an inverse
temperature: $T=\beta^{-1}$.  (The energy is written as $E_N$ in
\cite{CDE}, but here we adopt the convention that quantities with and
without a circumflex refer respectively to the ring geometry and to the
interval geometry introduced below.)  Clincy et al.~obtained the
Euler-Lagrange equations for the minimizers of the free energy functional,
which here is equivalent to the large deviation functional (LDF), of the
rescaled particle densities $\rho_\al(x)$, in the limit $N \to \infty$ with
$N_\alpha/N\to1/3$ and $i/N\to x$, and showed that the uniform density
profiles $\rho_\alpha(x) =r_\alpha= 1/3$ are always a solution of these
equations, but that for temperatures $T$ below the critical temperature
$T_c=\beta_c^{-1}= (2 \pi\sqrt{3})^{-1}$ there is also a nonuniform
solution of the linearized equations which has a lower free energy than the
uniform state.  They interpreted this as a second order transition at $T_c$
from the uniform to the nonuniform state, and confirmed by numerical
simulations that for $T>T_c$ the system is in a single phase with
essentially no correlations (they are $O(1/N)$) between the locations of
particles of different species, while for $T < T_c$ there is segregation of
the different species.  They further argued that this transition will
persist (possibly becoming first order) for unequal densities. In this case
the stationary state is no longer an equilibrium state and is in fact
unknown; despite this, they were able to obtain the LDF to order $\beta^2$
for all densities (see also \cite{BDLvW}).

In the present work we consider the weakly asymmetric ABC system on a
one-dimensional lattice of sites $i = 1,...,N$ with zero flux boundary
conditions: the dynamics are the same as above, except that a particle at
site $i=1$ (respectively $i=N$) can only jump to the right (respectively
left).  We shall refer to this geometry as an interval.  In contrast to the
situation on the ring, there is for this system {\it always}, whatever the
values of $N_A$, $N_B$, and $N_C$, an energy function $E_N$ (see
\eqref{eq:2.2} below) such that the dynamics satisfies detailed balance
with respect to an equilibrium measure $\nu_\beta=Z^{-1}\exp\{-\beta E_N\}$.

When $N_A=N_B=N_C=N/3$ the energy $E_N$ for the system on the interval
agrees with the energy $\hat E_N$ on the ring \cite{CDE}, in the sense that
if we (mentally) connect site $N$ to site 1 clockwise, and thus identify
each configuration $\zetau$ in the interval with a corresponding
configuration $\hat\zetau$ on the ring, then $E_N(\zetau)$ and
$\hat E_N(\hat\zetau)$ agree up to a constant which depends only on the
$N_\alpha$.  Thus the probabilities of the configurations $\zetau$ and
$\hat\zetau$ (at the same $\beta$) are the same, and the invariance under
rotations of $\hat E_N$ implies a rather surprising ``rotation'' invariance
of $E_N$ and of the Gibbs measure on the interval.  It follows also that
all the results obtained in the present paper for the interval model in the
case of equal densities give corresponding results on the ring.  When the
$N_\alpha$ are not all equal one might of course use $E_N$ similarly to
define an energy of a ring configuration, but this energy would then depend
on which site is chosen for the origin.  The resulting Gibbs measure on the
ring would be neither rotationally invariant nor invariant under the time
evolution of the ABC dynamics.

Using the Gibbsian nature of the invariant measure and following \cite{CDE}
we can obtain the free energy functional $\F(\rho_A,\rho_B,\rho_C)$ of the
density profiles $\rho_\alpha(x)$ in the scaling limit $N\to\infty$,
$i/N\to x$.  The parameters of the model in this limit are the inverse
temperature $\beta$ and the mean densities
$r_\alpha=\int_0^1\rho_\alpha\,dx$.  Our main goal is to determine, for
given values of these parameters, whether $\F$ has a unique minimizing
profile or whether, conversely, there is more than one minimizer and thus
coexistence of phases, and in either case the form of the minimizing
profile(s).  We are able to establish uniqueness at high temperatures by a
direct study of the functional $\F$.  We further show that minimizers must
satisfy the Euler-Lagrange equations for $\F$, and by a study of the
solutions of these equations we establish the nonuniqueness of minimizers
when $r_A=r_B=r_C=1/3$ and $T<T_c$; the nonuniqueness corresponds to the
lack of rotational invariance for the minimizers on the ring.  Conversely,
we establish uniqueness of the minimizers (on the interval) whenever the
densities are not all equal.  In the process we also obtain the form of the
minimizing density profiles as elliptic functions corresponding to pieces
of the periodic trajectory of a particle moving in a quartic confining
potential.

We remark that work on the ABC model on the ring has been carried out
recently by Bodineau et al.~\cite{BDLvW}.  They investigated analytically
the pair correlations, and thus also the local density fluctuations, in the
uniform state, and showed that the latter diverge as $T\searrow T_c$.  They
also suggested a possible formula for the large deviation and discussed the
possibility of deriving this from the macroscopic fluctuation theory
\cite{BDGJ}.  The model on the ring was also studied, in a different
context, by Fayolle and Furtlehner \cite{FF}, who obtained results which
agree, for the case in which the system on the ring has the same stationary
Gibbs measure as does the system on the interval, with those obtained here.
Results concerning the relation between the hydrodynamic equations for the
ABC model and the LDF were obtained by Bertini et al.~\cite{BDGJ}.

The outline of the rest of the paper is as follows.  In
Section~\ref{sec:model} we define the Gibbs measure with respect to which
the dynamics satisfy detailed balance.  We then study some properties of
this measure; in particular, we describe its ground states, which are
nonunique whenever two species have equal mean densities which are greater
than or equal to $1/3$.  We investigate also some properties of the
microscopic correlation functions at finite-temperature: the mean field
nature of the interactions in this system leads in the $N\to \infty$ limit
to local measures which are exchangeable.

In Section~\ref{sec:scaling} we consider the scaling limit of the model
and briefly discuss the nature of possible limiting density profiles. In
Section~\ref{sec:freeenergy} we obtain the Helmholtz free energy $\F$ as a
functional of the scaled densities ${\rho_\alpha(x), \ x\in [0,1]}$ (this
is equivalent to obtaining the LDF) as well as the Euler-Lagrange equations
satisfied by minimizers of $\F$; the proof that minimizers exist and must
satisfy these equations is postponed until Section~\ref{sec:existence}.  In
Section~\ref{sec:minimizers} we investigate the minimizers of $\F$ by
studying in detail the solutions of the Euler-Lagrange equations.  These
solutions, which describe all stationary points of $\F$, are given by
periodic (elliptic) functions describing the motion of a particle in a one
dimensional quartic potential.  There are many such solutions for large
$\beta$; despite this, we prove uniqueness of the globally minimizing
density profiles for all $\beta$ so long as the mean densities are not all
equal.  Details of the uniqueness proof are given in
Sections~\ref{sec:proofunique} and \ref{sec:typen}. In
Section~\ref{sec:phased} we discuss the phase diagram of the model and
describe a perturbation expansion around the uniform state.  In
Section~\ref{sec:convex} we show that $\F$ is a convex functional of the
$\rho_\alpha(x)$'s (with fixed $r_\alpha$'s) at high temperature, which
gives an alternate proof of the uniqueness of minimizers of $\F$ in this
regime.  A concluding discussion is given in Section~\ref{sec:conclusion}.

\section{The steady state of the  model}\label{sec:model}

A configuration $\zetau$ of the ABC model on the interval is an $N$-tuple
$(\zeta_1,\ldots,\zeta_N)$, with $\zeta_i=A$, $B$, or $C$.  We will let
$\eta_\alpha(i)$ be a random variable which specifies whether a particle of
species $\alpha$ is present at site $i$: $\eta_\alpha(i)=1$ if
$\zeta_i=\alpha$ and $\eta_\alpha(i)=0$ otherwise, so that
$\eta_A(i) + \eta_B(i) + \eta_C(i) = 1$ and
$N_\alpha=\sum^N_{i=1} \eta_{\alpha}(i)$.  The time invariant measure for
the weakly asymmetric dynamics described in Section~\ref{sec:intro} is a
canonical Gibbs measure,
\be
\nu_\beta(\zetau) = Z^{-1} \exp [-\beta
E_N(\zetau)], \label{eq:2.2}
\ee
where
\be
E_N(\zetau)  = {\frac{1}{N}} \sum^{N-1}_{i=1}
\sum^{N}_{j=i+1} [ \eta_C (i) \eta_B(j) + \eta_A (i) \eta_C(j) +
\eta_B(i) \eta_A (j)] \label{eq:2.3}
\ee
and $Z$ is the usual equilibrium normalization factor, that is, the
(canonical) partition function with fixed particle numbers $N_\alpha$.  To
verify this invariance one checks detailed balance for the dynamics: if
$\zetau$ is a configuration with particles of types $\alpha$ and $\alpha+1$
on sites $i$ and $i+1$, respectively (recall the convention that $\alpha$,
$\alpha+1$, \dots run cyclically through $A$, $B$, and $C$), and
$\zetau^{i,i+1}$ is the configuration with these particles interchanged,
then the transition rates are that $\zetau\to\zetau^{i,i+1}$ at rate
$e^{-\beta/N}$ and $\zetau^{i,i+1}\to\zetau$ at rate $1$, and the detailed
balance condition
$e^{-\beta/N}\cdot\nu_\beta(\zetau)=1\cdot\nu_\beta(\zetau^{i,i+1})$
follows from \eqref{eq:2.2} and \eqref{eq:2.3}.  For $\beta < \infty$,
$\nu_\beta$ is the unique stationary measure; this follows from the
transitivity of the dynamics on the set of all $N!/(N_A!N_B!N_C!)$
configurations consistent with $\sum_i \eta_\alpha (i) = N_\alpha$.  When
$\beta = 0$ all these configurations are equally likely.

One can rewrite (\ref{eq:2.3}) in various forms which differ from each
other only by functions of $N_A$, $N_B$, and $N_C$, and this does not
affect $\nu_\beta$ when the $N_\alpha$ are fixed.  As an example, an
equivalent energy is
 \be
E_N^*=\frac{1}{N}\biggl\{\, \sum^N_{j=1} j [\eta_B(j) - \eta_A(j)]
+ 3\sum^{N-1}_{i=1}\sum^N_{j=i+1} \eta_B(i)  \eta_A(j)\biggr\}.
\label{eq:2.4}
\ee
We shall generally use the form $E_N (\zetau)$ given in (\ref{eq:2.3}), as
it clearly exhibits the cyclic symmetry between the different species.  

\begin{rem}\label{WASEP} (a) The energy $E_N^*$ of \eqref{eq:2.4} is of
particular interest when the $B$ species is not present, i.e., when
$N_B=0$.  In this case we may call the $A$ particles just particles and the
$C$ particles holes, and the model reduces to the weakly asymmetric simple
exclusion process (WASEP).  Then $E_N^*=-N^{-1}\sum_jj\eta_A(j)$, which is
the energy arising from an external field of magnitude $N^{-1}$ pushing the
particles to the right.  See \cite{SS} and Section~7.2.2 of the review
\cite{BE} for discussions of the partially asymmetric simple exclusion
process on an interval.

 \smallskip\noindent
 (b) On the ring, the stationary measure for the WASEP gives equal weight
to all $\binom{N}{N_A}$  configurations.  In contrast
to the situation on the interval, however, the dynamics is not reversible,
i.e., does not satisfy detailed balance, with respect to this stationary
measure.  We thus have a true nonequilibrium stationary state (NESS).
\end{rem}

\subsection{Ground states}\label{sec:gs}
\label{sec:groundstate}

For $\beta =\infty$ there are many configurations stationary for the
dynamics: all those of the form $\bf\cdots ABCAB\cdots$, where a boldface
letter denotes a block of particles of the corresponding species.  These
configurations are local minimizers of the energy $E_N(\zetau)$, in the
sense that any interchange of adjacent particles of different species will
raise the energy, but not all of them are obtained as limits of $\nu_\beta$
when $\beta \to \infty$.  In this limit the measure will be concentrated on
the ground states of $E_N$, that is, on the global minimizers of
$E_N(\zetau)$ \cite{AL}.  As we will see, the energy per particle in the
ground state $\zetau$ is
$e_0=N^{-1}E_N(\zetau)=\min\{r_Ar_B,r_Ar_C,r_Br_C\}$, where
$r_\alpha\equiv N_\alpha/N$.

We describe the ground states by considering several cases. It is helpful
to note that \eqref{eq:2.3} may be summarized as saying that there is a
contribution of $1/N$ to the energy each time a $B$ particle lies to the
left of an $A$ particle, a $C$ to the left of a $B$, or an $A$ to the left
of a $C$. We also remark that if a local minimizer contains a sequence
$\bf BCAB$ of four blocks with a total of $k_\alpha$ particles of type
$\alpha$, then one may lower the energy by regrouping these particles into
three blocks, $\bf BCA$, $\bf ABC$, or $\bf CAB$, unless $k_A=k_C\ge k_B$
(and similarly for $\bf ABCA$ and $\bf CABC$).

 \smallskip\noindent
 (i) If one species is not present, say $N_\alpha=0$, then the ground state
energy is zero and there is a unique ground state configuration with all
particles of type $\alpha+1$ to the left of all particles of type
$\alpha+2$.  This is the WASEP discussed in \crem{WASEP}.  In this case
there are no local minimizers other than the global minimizer.  

 \smallskip\noindent
 From now on we suppose that none of the $N_\alpha$ vanish.  

 \smallskip\noindent
 (ii) If one of 
the $N_\alpha$ is greater than the other two, say $N_B>\max\{N_A, N_C\}$
(other cases can then be found by cyclic permutation), then  it is easy to
see that there is a unique ground state $\zetau$ consisting of three blocks
ordered as $\bf ABC$, with $e_0=r_A r_C$.

 \smallskip\noindent
 (iii) If two of the $N_\alpha$ are equal and the third is smaller, say
$N_A = N_C > N_B$, the ground state will be $N_B+1$ fold degenerate,
consisting of three or four blocks ordered as $\bf BCA$, $\bf CAB$, or
$\bf BCAB$, that is, some of $B$'s will be at the left side of the
interval, followed by all the $C$'s, then all the $A$'s, and finally the
remaining $B$'s.  Here $e_0=r_Ar_B=r_Br_C$.

 \smallskip\noindent
 (iv) If all of the $N_\alpha$ are equal, $N_A =N_B =N_C$, then the ground
states are the shifts, by an arbitrary number of sites, of the
configuration with three blocks ordered as $\bf ABC$; see the discussion in
Section~\ref{sec:intro} of the correspondence in this case with the model
on the ring.  The ground state is thus $N$-fold degenerate, and $e_0=1/9$.

\begin{figure}
\centerline{\includegraphics[width=6cm,height=6cm]{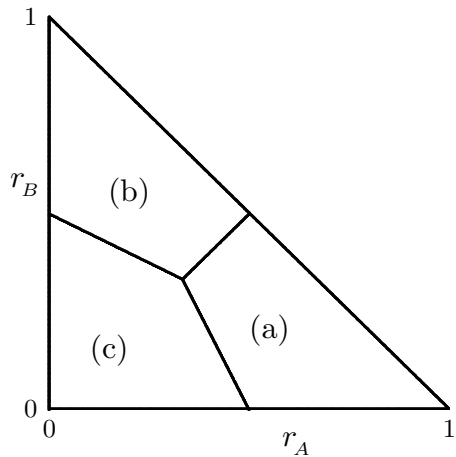}}
\caption{The $(r_A,r_B)$ ground state phase diagram.
It exhibits three phases separated by three first order
lines. In region (a), the A density is larger than that of the B's
and C's and hence the order is such that the A-block is in the
middle. Similarly (b) and (c)
correspond to regions where  B and C densities are largest.}
\label{fig:gsphase}
\end{figure}

The ground state phase diagram of the system in the right triangle where
$r_A,r_B\ge0$ and $r_A+r_B\le1$ is given in Figure~\ref{fig:gsphase}
(remember that $r_C = 1 -r_A - r_B$).  The ground state is unique
everywhere except on the three line segments originating from
$r_A=r_B= 1/3$ and terminating at the midpoints of the sides of the
triangle, corresponding to cases (iii) and (iv) above.  On the edges of the
triangle the system reduces to the WASEP.

\subsection{Finite temperature correlations and local
states}\label{sec:correlations}

The correlation functions at finite $\beta$  are 
 \be
   \langle \eta_{\alpha_1}(i_1)\eta_{\alpha_2}(i_2)\cdots
\eta_{\alpha_n}(i_n)\rangle,  \label{eq: 3.1}
 \ee
 where $\langle\cdot\rangle$ denotes the expectation with respect to the
Gibbs measure $\nu_\beta$.  We point out here some simple relations that
these satisfy, both in finite volume and in the limit $N\to\infty$,
$N_\alpha/N\to r_\alpha$.  (With some abuse of notation we will use
$r_\alpha$ to denote both the density $N_\alpha/N$ in a finite system and
the limiting value of the density in the $N\to\infty$  limit.)

First, it is easy to check from \eqref{eq:2.2} and
\eqref{eq:2.3} that, whenever all the $N_\alpha$ are positive,
\be
   \frac{\langle\eta_\alpha(N)\rangle}{\langle\eta_\alpha(1)\rangle} =
    e^{\beta[r_{\alpha+2} - r_{\alpha+1} ]},
\label{eq:3.2}
\ee
and this implies that for all $N$,
 \be
\prod^3_{\alpha = 1} \langle\eta_\alpha(N)\rangle
   = \prod^3_{\alpha=1} \langle\eta_\alpha(1)\rangle. 
\label{eq:3.3}
 \ee
 Note that when $r=(1/3,1/3,1/3)$ and the system on the interval coincides
with that on the ring, \eqref{eq:3.3} is a consequence of the translation
invariance of the latter.

 Next we note that there is a constant $K$, which depends only on $\beta$
and the $r_\alpha$, such that for fixed $x\in(0,1)$, $j_1$ and $j_2$
integers with $0<j_1<j_2$, and $\alpha_0$, $\alpha_1$, and $\alpha_2$
distinct particle species,
 \be\label{firstK}
\lim_{N\to \infty} \langle \eta_{\alpha_0} (\floor{xN}) 
  \eta_{\alpha_1}(\floor{xN}+j_1)\eta_{\alpha_2}(\floor{xN}+j_2)\rangle = K,
 \ee
 where $\floor{xN}$ denotes the largest integer less or equal to
$xN$.  To see that  the limit in  \eqref{firstK} is independent of $x$, $j_1$,
$j_2$, $\alpha_0$, $\alpha_1$, and $\alpha_2$ we first write 
$i_0=\floor{xN}$, 
$i_1=\floor{xN}+j_1$, and 
$i_2=\floor{xN}+j_2$, and observe that 
if $\zetau$ is a configuration with
$\zeta_{i_0}=\alpha_0$, 
$\zeta_{i_1}=\alpha_1$, and 
$\zeta_{i_2}=\alpha_2$, 
and $\underline\xi$ is the
configuration on $N-3$ sites
which is obtained from $\zetau$ by omitting the sites $i_1$, $i_2$, and
$i_3$   and
renumbering, then
$E_N(\zetau)=E_{N-3}(\underline\xi)+1+\Delta$; here the terms $1+\Delta$
arise from those terms in the energy \eqref{eq:2.3} involving the particles
on sites $i_1$, $i_2$, and $i_3$, and 
$\Delta$, which depends on $\xi$, the $i_k$, and the $\alpha_k$, satisfies 
$|\Delta|\le j_2/N$.  Thus from \eqref{eq:2.2} and \eqref{eq:2.3},
 \be
Z^{-1}\langle \eta_{\alpha_0} (i_0) 
  \eta_{\alpha_1}(i_1)\eta_{\alpha_2}(i_2)\rangle
 = \sum_\xi e^{-\beta E_{N-3}(\underline\xi)+1+\Delta}. \label{eq:Zcorr}
 \ee
 The corresponding expression for a different choice of the $i_k$ and
$\alpha_k$ will differ from \eqref{eq:Zcorr} only in the replacement of
$\Delta$ by some $\Delta'$, from which the result follows.

The results \eqref{eq:3.3} and \eqref{firstK} are closely related: each
shows that a product of three occupation numbers, or their expectations, is
site independent.  We will see another such relation in
Section~\ref{sec:minimizers}: for all scaled density profiles 
$\rho_\alpha(x)$ which are
stationary points of the free energy functional 
(see Section~\ref{sec:freeenergy}) the product
$\rA(x)\rB(x)\rC(x)$ is in fact independent of $x$.

Similar identities can help us understand the nature of local states
\cite{Ligg} in this model.  By the {\it local state} $\mu_{(x)}$ at position
$x\in(0,1)$ we mean the infinite volume state, if it exists, whose
correlation functions for particles of type $\alpha_1,\ldots,\alpha_h$ at
sites $\floor{xN}+j_1,\ldots,\floor{xN}+j_h$ are given by 
 \be\label{localstate}
 \lim_{N\to\infty} \Bigl\langle \prod^h_{i=1} 
   \eta_{\alpha_i} (\floor{xN} + j_i)\Bigr\rangle.
 \ee
 (If such a local state does not exist then one may identify
subsequences $N_k$ along which all the limits \eqref{localstate} do exist,
and thus obtain a family of local states.)  An argument similar to that
establishing \eqref{firstK} shows that for $x\in(0,1)$, $j_1,\ldots,j_h$
fixed, and $P$ any permutation of $\{1,\ldots,h\}$,
 \be
\lim_{N\to\infty} \frac{\Bigl\langle \prod^h_{i=1} 
   \eta_{\alpha_i} (\floor{xN} + j_i)\Bigr\rangle}
  {\Bigl\langle\prod^h_{i=1} 
  \eta_{\alpha_{P(i)}} (\floor{xN} + j_i)\Bigr\rangle} = 1.
\label{eq:3.5}
 \ee
 This implies \cite{HS} that $\mu_{(x)}$ is a product measure or a
superposition of product measures, which of course reflects the mean field
nature of the interaction.  Dynamically this corresponds to the fact that
when $N\to \infty$, the exchanges become symmetric and the invariant
measures are just superpositions of extremal product measures
\cite{Ligg}.  (In fact this last argument applies also on the ring with
unequal densities $r_\alpha$.)

\section{The scaling limit \label{sec:scaling}}

We now turn to the properties of the model in the macroscopic scaling
limit
 \be
N\to\infty, \quad N_\alpha/N\to r_\alpha, \quad i/N\to x.\label{hslimit}
 \ee
 We say that there is a unique (macroscopic) density profile in this limit
when there exists a function $\rho(x)=(\rho_A(x),\rho_B(x),\rho_C(x))$ such
that \cite{WC}
 \be\label{eq:2.6newa}
\lim_{N\to \infty}\Prob\left\{\left|
    \frac{1}{N} \sum_{i=1}^N\eta_\alpha(i)
    \varphi_\alpha\left(\frac{i}{N}\right)
    - \int^1_0 \rho_\alpha(x) \varphi_\alpha(x)\,dx\right|
 >\delta\right\}=0 ,
 \ee
 for all $\alpha$, all $\delta>0$, and all piecewise smooth functions
$\varphi_\alpha(x)$ on the interval [0,1].  Such $\rho_\alpha(x)$ must, of
course, satisfy the conditions
 \be\begin{gathered}
0\le\rho_\alpha(x)\le1\quad\hbox{and}\quad 
  \int^1_0 \rho_\alpha(x) dx = r_\alpha\quad\hbox{for}\quad \alpha=A,B,C,\\
 \sum_\alpha \rho_\alpha(x) = 1.\end{gathered}
   \label{eq:3.6}
 \ee

 Equation \eqref{eq:2.6newa} expresses the convergence, in a certain sense,
of the random profile $N^{-1}\sum_i\eta_\alpha(i)\delta_{i/N}$ to the
non-random profile $\rho_\alpha(x)$: specifically, for each
$\varphi_\alpha$ the random variable
$N^{-1}\sum_i\eta_\alpha(i)\varphi_\alpha(i/N)$ converges in the sense of
\eqref{eq:2.6newa} to the non-random quantity
$\int\rho_\alpha(x)\varphi_\alpha(x)\,dx$.  It may happen, on
the other hand, that the limiting profile is itself random, in the sense
that there is some family of limiting profiles
$\rho(\omega)=\{\rho_\alpha(\omega;x)\}$, indexed by the variable $\omega$
lying in some sample space $\Omega$ and each satisfying \eqref{eq:3.6}, and
a measure $\kappa$ on $\Omega$, such that for each test function
$\varphi_\alpha(x)$,
 \be\label{eq:superpos}
\lim_{N\to \infty}\frac{1}{N} \sum_{i=1}\eta_\alpha(i) \varphi_\alpha
\left(\frac{i}{N}\right) 
  = \int^1_0 \rho_\alpha(\cdot;x) \varphi_\alpha(x)\,dx
 \ee
 in the sense of convergence in distribution.  (In fact, \eqref{eq:2.6newa}
is also convergence in distribution, to a non-random limit.)  We would
then say that the limiting profile is not unique and that the system can
exist in more than one phase, i.e., is in a phase transition region.  This
is exactly what happens to the ferromagnetic Ising model in two or more
dimensions in a periodic box at temperatures below the critical
temperatures and fixed magnetization $m\in(-m^*,m^*)$, where $m^*$ is the
spontaneous magnetization \cite{Ising}.  In our system this happens, as we
shall see, if and only if $r_A=r_B=r_C$ and $T<(2\pi\sqrt3)^{-1}$.  More
generally, it may be that the limit in \eqref{eq:superpos} will not exist;
in that case, one would expect to obtain a limit in the given form by
passing to a subsequence.

\section{The free energy}\label{sec:freeenergy}

The question of whether or not there exists a limiting profile(s) as given
by \eqref{eq:2.6newa} or \eqref{eq:superpos}, and of which of these
alternatives applies, is related to the question of the existence and
uniqueness of minimizers of the LDF for this system \cite{LD}.  If
$n(x)=(n_\alpha(x))_{\alpha=A,B,C}$ denotes a general continuum density
profile satisfying \eqref{eq:3.6} (written in terms of $n$ rather than
$\rho$) then the logarithm of the probability of finding 
the profile $n(x)$ in the scaling
limit is asymptotically $-N\F_{\rm LD}(\{n(x)\}$, where $\F_{\rm LD}$
is the LDF.  Possible limiting density profiles are thus those which
minimize $\F_{\rm LD}$.

To obtain from the microscopic measure the probability of finding the
profile $n(x)$ it is necessary (roughly speaking) to sum
$\nu_\beta(\zetau)$, as given in \eqref{eq:2.2}, over all (microscopic)
configurations $\zetau$ consistent with $n(x)$;  this summation
will yield a restricted partition function divided by the full partition
function $Z$.  Taking the logarithm of this ratio, dividing by $N$, and
taking the scaling limit of the result will then yield
 \be
\F_{\rm LD}(\{n(x)\})=\F(\{n(x)\}) - \inf_{n(x)}\F(\{n(x)\}),
 \ee
  where $\beta^{-1}\F(\{n(x)\})$ is the Helmholtz free energy of the system
  restricted to having a density profile $n(x)$:
 \be
\F(\{n(x)\})=\beta \E(\{n(x)\})-\S(\{n(x)\}).
 \ee
 Here the macroscopic energy $\E(\{n(x)\})$ and entropy $\S(\{n(x)\})$ are
the limits of the microscopic energy and entropy per site; as indicated,
these limits depend only on the profile $n(x)$.  For the energy, this is
due to the mean field nature of the microscopic energy $E_N$ of
\eqref{eq:2.3}.  For the entropy, which is the logarithm of the number of
microscopic configurations consistent with $n(x)$, this is due to the
Bernoulli nature of each component of the local measure, which implies that
all microscopic configurations consistent with $n(x)$ have the same energy
and hence the same weight.  Thus
 \bea
\F\left( \{n\}\right) &=& \beta \int_0^1 \,dx
\int_x^1\,dz  \left[n_{A}(x)n_C(z)+ n_B(x)n_{A}(z) + n_C(x)n_B(z)\right] \cr
&&\hskip-35pt+ \int_0^1 dx \left[n_{A}(x) \ln n_{A}(x)+n_B(x) \ln n_B(x)
    +n_C(x) \ln n_C(x)\right]\cr
&&\hskip-45pt=\; \sum_\alpha\int_0^1\left[\beta \int_0^1 
\Theta(z-x)n_\alpha(x)n_{\alpha+2}(z)\,dz
 +n_{\alpha}(x) \ln n_{\alpha}(x)\right]\,dx. 
\label{eq:F}
 \eea
 The profiles $n(x)$ minimizing \eqref{eq:F}---that is, the candidates for
limiting density profiles---will represent a compromise between the
entropy, which wants to keep all the densities uniform, and the energy,
which wants to keep the different particle types segregated.

The next theorem establishes the existence and properties of these
minimizers.
 
\begin{thm}\label{thm:exist} Suppose that $0<\beta<\infty$.  Then given
positive numbers $r_A$, $r_B$, and $r_C$ with $r_A + r_B +r_C =1$, let
$F(r_A,r_B,r_C)$ denote the infimum the values of
${\cal F}(\{n\})$ over (measurable) profiles $n(x)$ satisfying the
constraints \eqref{eq:3.6}.  Then there exist infinitely differentiable
functions $\rho_A(x)$, $\rho_B(x)$ and $\rho_C(x)$, possibly non-unique,
that satisfy these constraints and achieve the minimum:
 \be 
  {\cal  F}(\{\rho\}) = F(r_A,r_B,r_C)\ .
 \ee
Moreover, there is a $\delta>0$ so that each $\rho_\alpha$ satisfies
$\delta < \rho_\alpha(x) < 1-\delta$ for all $x$; thus the minimizer is an
interior point with respect to the constraint of taking values in $[0,1]$
and so satisfies the Euler-Lagrange equations (ELE) obtained from $\F$.
\end{thm}

The proof of this theorem is given in Section~\ref{sec:existence}.  We note
here, however, that the minimizers lie in the interior of the constraint
region essentially because the entropy term has infinite normal derivative
at the boundary.  For $\beta= \infty$, when only the energy counts, the
densities minimizing $\lim_{\beta\to\infty}\beta^{-1}\F=\E$ will be the
continuum limit of the ground state configurations described in section
\ref{sec:groundstate} and need not satisfy the ELE.

To obtain the Euler-Lagrange equations satisfied by the minimizing
profiles we must take the variational derivatives of $\F\left(\{n\}\right)$
with respect to two of the density profiles while maintaining
$\sum_\alpha n_\alpha=1$ and $\int_0^1n_\alpha(x)\,dx=r_\alpha$; here we
will treat $n_A(x)$ and $n_B(x)$ as independent, with
$n_C(x)=1-n_a(x)-n_B(x)$.  Defining
$\F_\alpha(x)=\delta\F/\delta n_\alpha(x)$ to be the variational derivative
taken as if the profiles $n_A(x)$, $n_B(x)$, and $n_C(x)$ were independent,
we will then have
 \be
\left.\frac{\delta\F}{\delta n_A}\right|_{n_C=1-n_A-n_B}=\F_A-\F_C,
   \quad
\left.\frac{\delta\F}{\delta n_B}\right|_{n_C=1-n_A-n_B}=\F_B-\F_C.
 \ee
 Imposing the constraints $\int_0^1 n_\alpha(x)\,dx=r_\alpha$ leads to the
ELE
 \be\label{ELE0}
 \F_A-\F_C={\rm constant}, \qquad  \F_B-\F_C={\rm constant},
 \ee
 where
\be\label{eq:FA}
\F_\alpha(x)= \log n_\alpha(x) 
   + \beta \int_0^x[n_{\alpha+1}(z)-n_{\alpha+2}(z)]\,dz+1+\beta r_{\alpha+2}.
\ee
Simple manipulations then show that under the constraint
$\sum_\alpha n_\alpha(x)=1$ the derivatives $\F_\alpha$ satisfy
 \be\label{eq:ident}
\sum_{\alpha=A,B,C} n_\alpha
  \frac{\partial}{\partial x}\F_\alpha = 0,
 \ee
 which implies, rather surprisingly, that solutions of
\eqref{ELE0} will satisfy $\F_\alpha(x)=\rm constant$ for all $\alpha$, i.e.,
that the functional derivatives of $\F$ can be taken as if the $n_\alpha$
were independent.  From \eqref{eq:FA}, then, the minimizing profile
$\rho(x)$ must satisfy the ELE
\be
 \rho_\alpha(x) 
   = \rho_\alpha(0) 
     e^{\beta\int^x_0 [\rho_{\alpha+2}(y)-\rho_{\alpha+1}(y)]dy},
 \label{eq:int}
\ee
 for all $x\in[0,1]$.  It follows that
\be
 \rho_\alpha(1) 
   =\rho_\alpha(0)e^{\beta(r_{\alpha+2}-r_{\alpha+1})},  \label{eq:int2}
\ee
 which is consistent with \eqref{eq:3.2}.  The ELE may also be written in
differential form:
  \begin{subequations}\label{eq:dABC}\begin{align}
 \frac{d\rA}{dx} &=  \beta\rA  \left(  \rC -\rB \right),\label{eq:dABCA}\\
 \frac{d\rB}{dx} &=  \beta \rB  \left(  \rA -\rC \right),\label{eq:dABCB}\\
 \frac{d\rC}{dx} &=  \beta \rC  \left(  \rB -\rA \right). 
\label{eq:dABCC}
 \end{align} \end{subequations}
 These equations were derived in \cite{CDE} for the case
$r=(1/3,1/3,1/3)$.  One obtains minimizing profiles by solving these
equations with the constraints
 \be\label{eq:ralpha}
 \int_0^1 \rho_\alpha(x)\,dx=r_\alpha. 
 \ee

\begin{rem} As indicated above, one may  consider $\F$ as a
function of $n_A$ and $n_B$ alone, with $n_C(x)=1-n_A(x)-n_B(x)$.  The
corresponding variational derivatives $\lambda_A(x)$ and $\lambda_B(x)$
are the local chemical potentials for the density profiles,
 \be
\lambda_A=\F_A-\F_C,\qquad \lambda_B=\F_B-\F_C.
 \ee
 which are constant at a stationary point of $\F$ (see \eqref{ELE0}).
These are given by
\bea\label{eq:4.4a}
\frac{\delta \F}{\delta n_A(x)} 
  &=& \lambda_A(\{n_A\},\{n_B\})\nonumber\\
  &=& \log\left[ \frac{n_A(x)}{n_C(x)}\right]
   - \beta \int_0^x(1-3n_B(z)) dz  +\beta(r_C-r_B),\\
\label{eq:4.4b}
\frac{\delta \F}{\delta n_B(x)} 
  &=&\lambda_B(\{n_A\},\{n_B\})\nonumber\\
  &=& \log\left[ \frac{n_B(x)}{n_C(x)}\right] 
  - \beta \int_0^x (1-3n_A(z)) dz + \beta(r_A-r_C),\quad  
\eea
 with the replacements $n_C(x)=1-n_A(x)-n_B(x)$ and $r_C=1-r_A-r_B$.
\end{rem}

\section{Solutions of the ELE equations}\label{sec:minimizers}

We now turn to a detailed study of the solutions of the Euler-Lagrange
equations \eqref{eq:dABC} with the constraints \eqref{eq:ralpha}.  As
already noted in \cthm{thm:exist}, for any given $\beta$ and
$r_A, r_B, r_C$ there must exist at least one solution of these,
corresponding to a minimizer of $\F$.  There may, however, be many
solutions; these will all correspond to stationary points of $\F$, and more
than one of these stationary points may be a minimizer.  In this section we
determine all the minimizers.

\subsection{The WASEP}\label{sec:WASEP}
 
We first consider briefly the case in which one of the densities is zero,
say $r_B=0$, when the system reduces to the WASEP (see \crem{WASEP}).  This
system, which can be solved explicitly on the interval, has in
the scaling limit a unique limiting density profile---i.e., minimizer of
the free energy---given by the solution of (\ref{eq:dABC}) with
$\rho_B(x) = 0$:
\be
\rho_A(x) = \frac{De^{\beta x}}{1+De^{\beta x}}=1-\rho_C(x), \qquad  x\in
[0,1]. 
\label{eq:4.7}
 \ee
 The constant $D$ is determined by the constraint
$\int^1_0\rho_A(x)dx=r_A=1-r_C$.  When $\beta\rightarrow \infty$,
 \eqref{eq:4.7} reduces to the ground state configuration in which  all the
$A$ particles are pushed to the right:
 \be
 \rho_A(x)\big|_{\beta=\infty} = \begin{cases}
   0,&\text{if $0\leq x\leq 1 - r_A$,} \\
   1,&\text{if $1- r_A \leq x\leq 1$.}\end{cases}
 \ee
 The local measure $\mu_{(x)}$ in this model is, as expected, the Bernoulli
measure with density $\rho_A(x)$.  (On the ring the stationary state of the
WASEP is a nonequilibrium one in which all configurations have equal weight
and the limiting scaled density profile is constant.)

\subsection{Properties of solutions.}\label{view} Through the remainder of
this section we suppose that all the $r_\alpha$ are strictly positive.  Let
us begin by discussing the properties of some given solution
$\rho(x)=(\rho_A(x),\rho_B(x),\rho_C(x))$ of the ELE equations
\eqref{eq:dABC} with the constraints \eqref{eq:ralpha}. By differentiating
$\log(\rho_A(x)\, \rho_B(x)\, \rho_C(x))$ with respect to $x$ we see that
the ELE imply that there is a constant $K$ such that
 \be\label{eq:product}
   \rho_A(x)\, \rho_B(x)\, \rho_C(x) = K,\qquad 0\le x\le1,
 \ee
  and  that the condition
 \be \label{eq:sum}
 \rho_A(x) + \rho_B(x) + \rho_C(x) =1, \qquad 0\le x\le1,
 \ee
 is preserved by the equations.  Equation \eqref{eq:product} is the
 scaling limit version of  \eqref{firstK}.
 
Since $\rho_A(x)$ and $\rho_C(x)$ have sum $1-\rho_B(x)$ and product
$K/\rho_B(x)$ they must be the two roots of the equation
$r^2+(1-\rho_B(x))r+K/\rho_B(x)=0$, from which 
 \be\label{diff}
\rho_A(x)-\rho_C(x)=\pm\sqrt{\frac{\rho_B(x)(1-\rho_B(x))^2-4K}{\rho_B(x)}}.
 \ee
 Then squaring \eqref{eq:dABCB} and using \eqref{diff} we find that $\rho_B(x)$
 is a solution of 
 \be\label{osc0}
 {\rho'(x)}^2+8\beta^2U_K(\rho(x))=0,
 \ee
 where 
 \be\label{eq:potential}
     U_K(\rho) \equiv \frac12K\rho-\frac18\rho^2(1-\rho)^2.
 \ee
 Of course, the same argument shows that $\rho_A(x)$ and $\rho_C(x)$ must
also be solutions of \eqref{osc0}.

\begin{figure}[ht!]
\begin{center}\includegraphics[width=10cm]{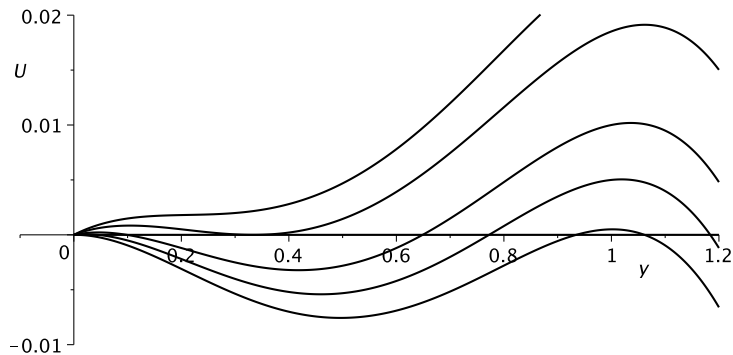}

\caption{Plots of $U_K(y)$ for $K=1/20$, $1/27$, $1/50$, $1/100$,
 and $1/1000$.} 
\label{pots}
\end{center}
\end{figure}

To study \eqref{osc0} we write $t=2\beta x$ and let $y(t)=\rho(t/2\beta)$;
then $y$ satisfies 
 \be\label{osc}
 \frac12{y'(t)}^2+U_K(y(t))=0.
 \ee
 This is the equation of the zero energy solution of a mass 1 particle
moving in a potential $U_K$.  Because $U_K$ is quartic in $y$ the solutions
are elliptic functions; see Appendix~\ref{sec:elliptic}.  For $K>1/27$,
$U_K(y)$ is strictly positive for $0<y<1$ and so no solutions with $y$ in
this range exist.  For $K=1/27$, $U_K(y)$ has a local minimum with value 0
at $y=1/3$, and \eqref{osc} has constant solution $y(t)=1/3$.  For
$0<K<1/27$, $U_K$ has four zeros, $0$, $a(K)$, $b(K)$, and $c(K)$, where
$0<a<b<1<c$ and $U_K(y)<0$ for $a<y<b$.  See Figure~\ref{pots}.  Since we
are interested in solutions of \eqref{eq:dABC} which satisfy
$0<\rho_\alpha(x)<1$ we consider only the solutions of \eqref{osc} which
oscillate between $a$ and $b$.  Let $y_K$ denote the solution of
\eqref{osc} which satisfies $y_K(0)=a$; $y_K(t)$ then has period
 \be\label{period}
\tau_K = 2\int_a^{b}\frac{dy}{\sqrt{-2U_K(y)}}\;.  
 \ee
 For $0\le t\le \tau_K/2$, $y_K(t)$ is determined by inverting the relation
 \be\label{solution}
t = t(y)= \int_a^{y}\frac{dw}{\sqrt{-2U_K(w)}}\;;  
 \ee
 $y_K(t)$ is then obtained for all $t$ by extending to an even function of
period $\tau_K$.

\begin{rem} \label{rem:trivial} We note some properties of this
  solution; those listed in (a) are elementary, and we relegate the proofs
  of (b) and (c) to Appendix~\ref{proofs}.

 \smallskip\noindent
 (a) $y_K(t)$ is even, is periodic with minimal period $\tau_K$, has local
minima at integer multiples of $\tau_K$ and local maxima at half integer
multiples of $\tau_K$, and is monotonic between these points.  Moreover,
$y_K(t)=y_K(s)$ if and only if  either $s+t$ or $s-t$ is an
integer multiple of $\tau_K$.

 \smallskip\noindent
 (b) $\lim_{K\nearrow1/27}\tau_K=4\pi\sqrt3=2\beta_c$ and
$\lim_{K\searrow0}(\tau_K/\ln(1/K))=6$. Moreover, for any $\epsilon>0$,
 \be\label{perints1}
 \lim_{K\to0}\frac1{\ln(1/K)}\int_a^\epsilon\frac{dy}{\sqrt{-2U_K(y)}}=2, 
 \ee
 and
 \be\label{perints2}
 \lim_{K\to0}\frac1{\ln(1/K)}\int_{b-\epsilon}^{b}\frac{dy}{\sqrt{-2U_K(y)}}
   =1.  
 \ee
This means that in the limit $K\to\infty$ the trajectory $y_K(t)$ will spend
two-thirds of its time near the origin and one-third near $y=1$; see for
example the third trajectory shown in Figure~\ref{fig:solns}.

 \smallskip\noindent
 (c) $\tau_K$ is a strictly monotonic decreasing function of $K$ for
$0<K\le1/27$.  \end{rem}

\begin{figure}[htb!]
\begin{center}
\includegraphics[width=12cm]{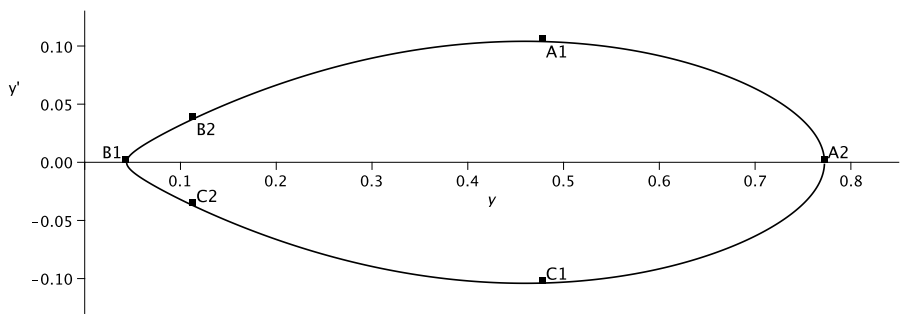}

\caption{Phase plane orbit of $y_K(t)$ for $K=1/100$.  The time intervals
between the six marked points are all equal.} \label{fig:orbit} 
\end{center}
\end{figure}

\begin{figure}[htb!] 
 \begin{center} 
\includegraphics[width=12cm, height=3.2cm]{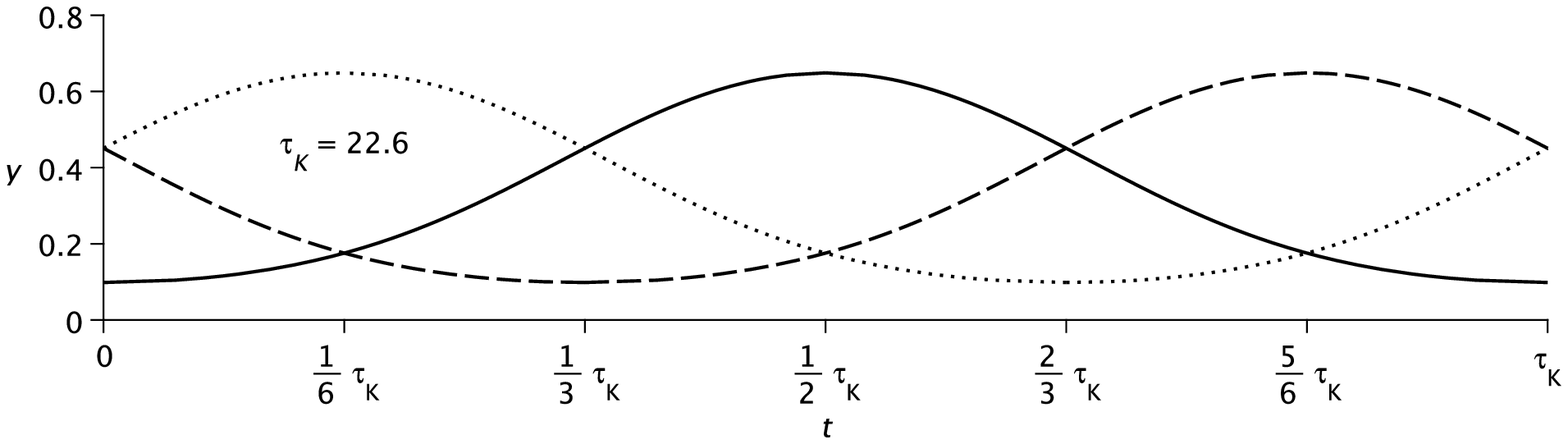}
\includegraphics[width=12cm, height=4cm]{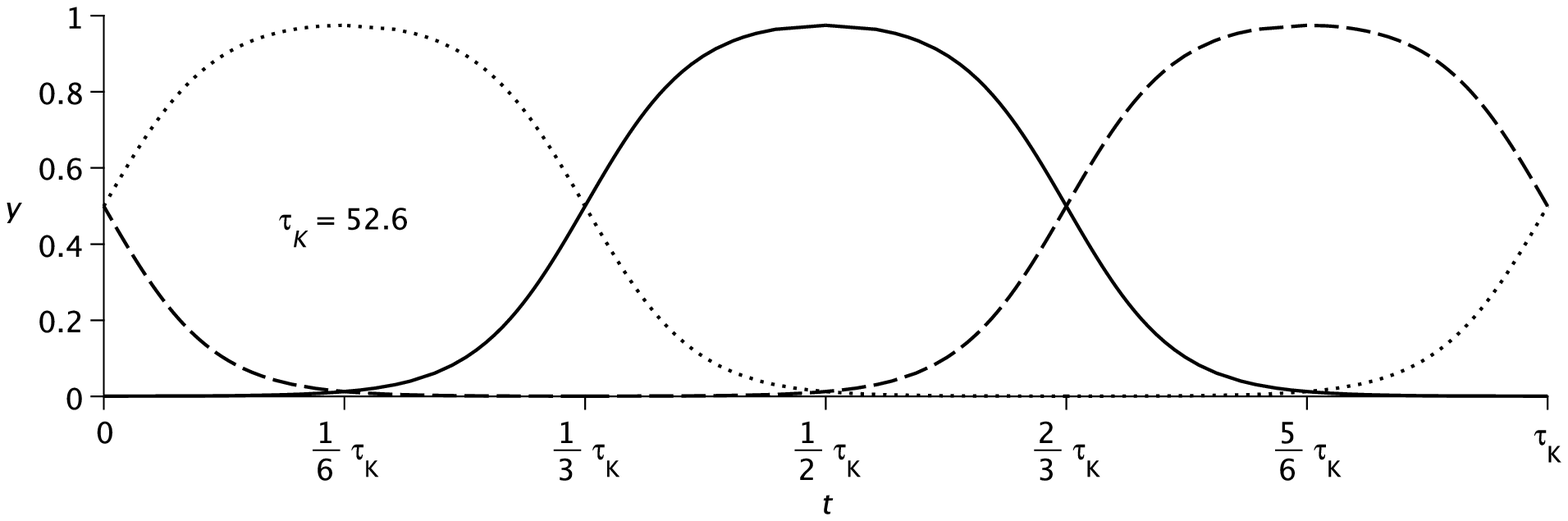}
\includegraphics[width=12cm, height=4cm]{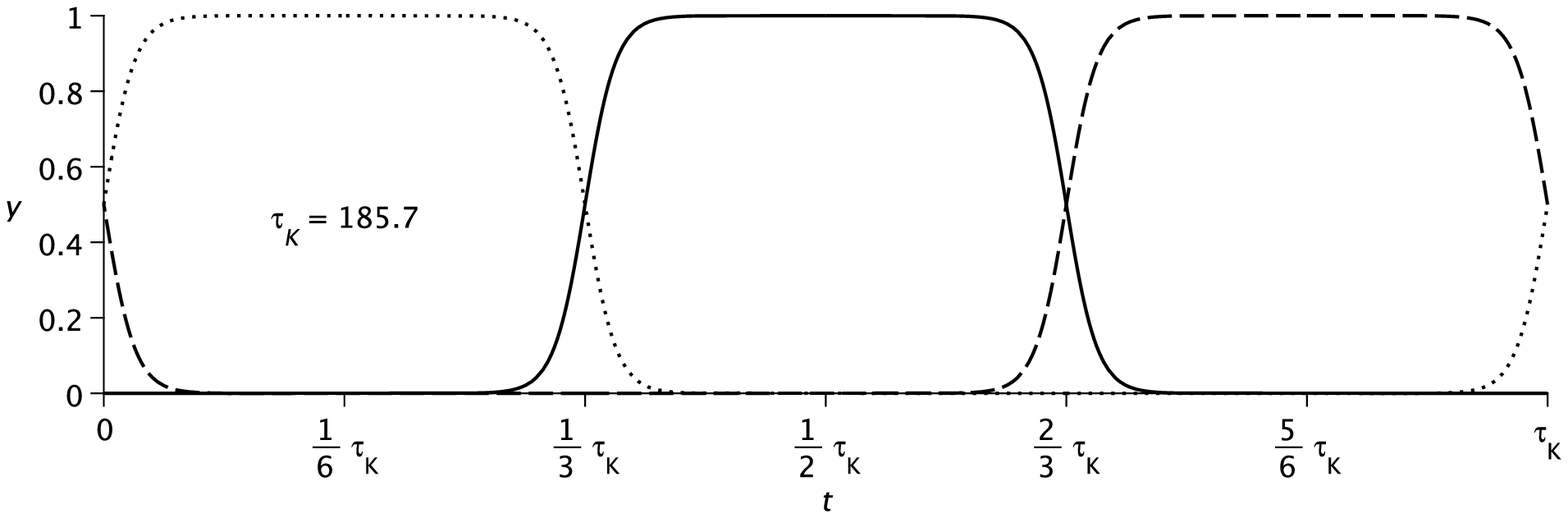}

\caption{Plots of $y_K(t)$ (solid),
$y_K(t+\tau_K/3)$ (dotted), and $y_K(t-\tau_K/3)$
(dashed) for $K=1/50$ (top), $K=1/6400$ (middle), and
$K=5^{-2}2^{-40}=1/27487790694400$ (bottom).} \label{fig:solns} 
\end{center}
\end{figure}

Now for $\alpha=A,B,C$ there must be a phase shift $t_\alpha$ such that 
 \be\label{solns}
\rho_\alpha(x)=y_K(2\beta(x-1/2) + t_\alpha), \qquad 0\le x\le 1,
 \ee
 that is, each $\rho_\alpha(x)$ is obtained by looking at the solution
$y_K(t)$ within a window of length $2\beta$ centered at some value
$t_\alpha$, and rescaling from $t$ to $x$. 
The phase shifts are not independent; in fact,
 \be\label{eq:phases}
   t_A=t_B+\tau_K/3 \qquad\hbox{and}\qquad t_C=t_B-\tau_K/3
 \ee
 (see (27) of \cite{CDE}). 
 In verifying \eqref{eq:phases} the requirement \eqref{eq:ralpha} of certain
average densities is not relevant.  What matters is that we consider three
solutions of the ELE \eqref{eq:dABC} or equivalently (with the rescaling
$t=2\beta x$) three solutions $y_{K,\alpha}(t)$ of \eqref{osc} satisfying
 \be\begin{split}\label{odey}
y_{K,A}'&=y_{K,A}(y_{K,C}-y_{K,B})/2, \\
y_{K,B}'&=y_{K,B}(y_{K,A}-y_{K,C})/2,\\
y_{K,C}'&=y_{K,C}(y_{K,B}-y_{K,A})/2,\end{split}
 \ee
 with $y_{K,\alpha}(t)=y_K(t+t_\alpha)$.  We may assume that these are
defined for all $t$ and  without loss of generality that $t_B=0$.  It
is helpful to view the trajectories in the $y$-$y'$ phase plane; see
\cfig{fig:orbit}.  At time $t=0$ the B trajectory lies at the point on the
$y$ axis marked B1.  Since the velocity on the B trajectory is zero at this
point, it follows from \eqref{odey} that $y_{K,A}(0)=y_{K,C}(0)$, that is,
the A and C trajectories lie at points such as those marked  A1 and
C1, respectively; again from \eqref{odey} we know that A1 must be in the
upper half of the phase plane, since $y_{K,A}'(0)>0$. Now follow the motion
for a time period $\Delta t$ until $y_{K,A}$ reaches the $y$ axis at point
A2; since now $y_{K,A}'(\Delta t)=0$ it follows that the B and C
trajectories are at points B2 and C2 with equal $y$ coordinates, and it
follows immediately (using the symmetry around the $y$ axis) that the
travel time along each of the six orbital segments in the figure is
$\Delta t$.  Thus $\Delta t=\tau_K/6$, $t_A=t_B+2\Delta t=t_B+\tau_K/3$, and
$t_C=t_B-\tau_K/3$.

Let us introduce the notation
 \be\label{eq:defav}
\langle z\rangle_I=\frac1{|I|}{\int_Iz(t)\,dt},
 \ee
 where $|I|=\int_I\,dt$, for the average of the function $z$ over the set
$I$ (which will always be a union of intervals), and define
 $Y_K(t)=\langle y_K\rangle_{[t-\beta,t+\beta]}$.  Then with
 \eqref{solns} and \eqref{eq:phases},  \eqref{eq:ralpha} becomes 
 \be\label{eq:ralphay} Y_K(t_B)=\rhobar_B, \quad Y_K(t_B+\tau_K/3)=\rhobar_A,
\quad Y_K(t_B-\tau_K/3)=\rhobar_C.
 \ee
 The problem of solving \eqref{eq:dABC} and \eqref{eq:ralpha} is now
the problem of finding $K$ and $t_B$ satisfying \eqref{eq:ralphay}.  If
either $K=1/27$ so that $y_K(t)=1/3$ is constant, or $2\beta$ is an integer
multiple of $\tau_K$, then  $Y_K(t)=1/3$ is constant and \eqref{eq:ralphay}
has a solution (with arbitrary $t_B$) if and only if $r_A=r_B=r_C=1/3$. 

 Figure~\ref{fig:solns} shows the curves $y_K(t)$ and $y_K(t\pm \tau_K/3)$
for several values of $K$.  To obtain a solution of \eqref{eq:ralphay} for
some $K$ one views the corresponding three curves in a window of length
$2\beta$ (corresponding to the full lattice, i.e., to the original unit
interval after the variable change in \eqref{solns}) centered at $t_B$.  In this context we label a
solution by an integer which is one more than the number of full periods
(plus, perhaps, a fraction of a period) fit into the window: we say that a
solution $\rho(x)$ of \eqref{eq:ralphay} with $K<1/27$ is of {\it
type~$n$}, for $n=1,2,\ldots$, if
 \be\label{eq:typedef}
(n-1)\tau_K<2\beta\le n\tau_K.
 \ee
   We do not assign a type to the constant solution $y_K=1/3$, which exists
for $K=1/27$, $r_A=r_B=r_C=1/3$, and every value of $\beta$, as discussed
above.

\subsection{Uniqueness of solutions}\label{sec:preunique} 

The next theorem summarizes our results about solutions of the ELE and
answers the questions posed in the first paragraph of this section.  We
consider the problem of finding $y_K$ and $t_B$ satisfying \eqref{eq:ralphay}.

 \begin{thm}\label{thm:main} (a) If $r_A=r_B=r_C=1/3$ then there exist
(i)~the constant solution, (ii)~for $\beta>n\beta_c=2\pi n\sqrt3$,
$n=1,2,\ldots$, a solution of type $n$ unique up to translation, and
(iii)~no other solutions.  The minimizer of the free energy is, for
$\beta\le\beta_c$, the (unique) constant solution and, for $\beta>\beta_c$,
the type~1 solution.

\smallskip\noindent
 (b) For values of $r$ other than
$(1/3,1/3,1/3)$ there exists for all $\beta$ a unique type~1 solution which
is a minimizer of the free energy.  \end{thm}

In the remainder of this section we prove part~(a) of this theorem and give
an overview of some parts of the proof of part~(b).  More technical
parts of the proof of (b) are given in Sections~\ref{sec:proofunique} and
\ref{sec:typen}.  Before beginning, we summarize some simple properties of
the function $Y_K$ of \eqref{eq:ralphay}.

\begin{rem} \label{rem:averages} (a) As remarked above, if either $K=1/27$
or $2\beta$ is a multiple of $\tau_K$ then $Y_K(t)=1/3$ for all $t$.

 \smallskip\noindent
 (b) For all $K,\beta$ other than those of (a) a plot $Y_K(t)$ and
$Y_K(t\pm\tau_K/3)$ will look much like one of the sets of curves in
Figure~\ref{fig:solns}.  In particular, since $y_K$ is even and
$\tau_K$-periodic, so is $Y_K$, and from
$Y_K'(t)=(2\beta)^{-1}[y_K(t+\beta)-y_K(t-\beta)]$ it follows that $Y_K$
is strictly monotonic between integer multiples of $\tau_K/2$.  From this we
see that  if $Y_K(t)=Y_K(s)$ then either $s+t$ or $s-t$ must be an
integer multiple of $\tau_K$.

 \smallskip\noindent
 (c) To find all triples $(r_A,r_B,r_C)$ for which \eqref{eq:ralphay} has a
solution for a given $K$ it suffices to consider values of $t_B$ satisfying
$0\le t_B\le \tau_K/6$.  For such $t_B$ the possible triples satisfy
$r_B\le r_C\le r_A$; other orderings of the same sets of values are found
for other ranges of $t_B$.
 \end{rem}
 
 \begin{proofof}{Theorem \ref{thm:main}(a)}  When
$\rhobar=(1/3,1/3,1/3)$, \eqref{eq:ralphay} becomes
 \be\label{eq:3thirds}
 Y_K(t_B)=Y_K(t_B+\tau_K/3)=Y_K(t_B-\tau_K/3)=1/3.
 \ee
  For every $\beta$, one solution of \eqref{eq:3thirds} is the constant
$y_K(t) (=\rho_\alpha(x))=1/3$ corresponding to $K=1/27$.  Moreover, if
$2\pi n\sqrt3<\beta\le 2\pi(n+1)\sqrt3$ then by Remark~\ref{rem:trivial}
there exist $n$ one-parameter families of nonconstant solutions (indexed by
$t_B$) obtained by solving
 \be\label{eqj}
  2\beta = j\tau_{K_j}, \qquad j=1\ldots,n,
 \ee
  for $K_j$.  Note that the solution \eqref{eqj} is of type $j$.  No other
solutions are possible, because \eqref{eq:3thirds} is inconsistent with the
final observation in \crem{rem:averages}(b).

To complete the proof we note that, as is easily seen, perturbing the
uniform solution via $\rho_\alpha(x)\to1/3+\epsilon\cos(2\pi (x+x_\alpha))$,
where $x_A=x_B+1/3$ and $x_C=x_B-1/3$, decreases the free energy when
$\beta>\beta_c$, so that this solution cannot be a minimizer, and that we
will prove in Section~\ref{sec:typen} (see \cthm{rearrange}) that no
solution of type~$n$ with $n\ge2$ can minimize the free energy.
\end{proofof}

\begin{proofof}{Theorem \ref{thm:main}(b)} It was shown in \cthm{thm:exist}
that for any positive $r_A$, $r_B$, and $r_C$
which satisfy $\sum_\alpha r_\alpha=1$ there exists at least one 
minimizer of the free energy $\F$, and that this minimizer satisfies the
ELE.  By \cthm{rearrange} no solution of type~$n$ with $n\ge2$ can minimize
the free energy, so this minimizer must be of type 1.  It remains to prove
uniqueness of this solution; we will give the full proof in
Section~\ref{sec:proofunique} (see \cthm{thm:unique}), but here we
illustrate the idea of the proof by sketching the argument for the special
case $r_B<1/3$, $r_A=r_C>1/3$, in which symmetry considerations
considerably simplify the discussion.

We suppose then that for some such $r$ and some $\beta$ there are two
distinct type 1 solutions of \eqref{eq:ralphay}, and derive
a contradiction.  It follows from $r_A=r_C$ and \crem{rem:averages}(b)
that the phase shift $t_B$ must be zero for each solution,
so that we are looking at some $z_1=y_{K_1}$ and $z_2=y_{K_2}$ on the
interval $J=[-\beta,\beta]$ which satisfy (see \eqref{eq:defav})
 \be\label{eq:same1}
 \langle z_1\rangle_J=\langle z_2\rangle_J=r_B.
 \ee
 We take $K_2<K_1$ and write $\theta_i=\tau_{K_i}/2$, so that (because these are
type~1 solutions) $\beta<\theta_1<\theta_2$.  The situation is as in
Figure~\ref{fig:symm};
\begin{figure}[htb!]\label{fig:symm}
\begin{center}\includegraphics[width=11cm]{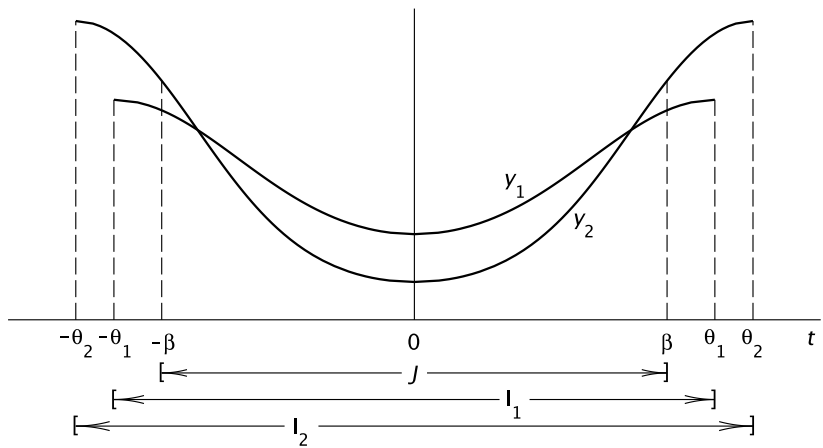}
\caption{Configurations of two solutions $z_1$, $z_2$ which are candidates
to satisfy $\langle z_1\rangle_J=\langle z_2\rangle_J=r_B$.}
 \end{center}\end{figure}
 the qualitative features of this figure, which we use below, are obtained
from $z_2(0)<z_1(0)$ (which holds because $y_K(0)=a(K)$, the smallest
positive zero of the potential $U_K(\rho)$ \eqref{eq:potential}, decreases
with $K$), the symmetry, and the fact that \eqref{eq:same1} implies that
the curves $z_1$ and $z_2$ must cross in the interval $J$.  In the full
proof below we will verify that the curves cross exactly twice, as shown.

 We write $I_1=[-\theta_1,\theta_1]$ and $I_2=[-\theta_2,\theta_2]$ and observe
 that
 \be\label{eq:third1}
\langle z_1\rangle_{I_1}=\langle z_2\rangle_{I_2}=1/3.
 \ee
 Since $\langle z_2\rangle_{I_2}$ is a (weighted) average of
$\langle z_2\rangle_{J}=r_B<1/3$ and $\langle z_2\rangle_{I_2\setminus J}$,
necessarily 
 \be\label{eq:ineq1}
\langle z_2\rangle_{I_2\setminus J}>r_B.
 \ee
  Moreover,
 \be\label{eq:ineq2}
   \langle z_2\rangle_{I_2\setminus J} 
   \ge \langle z_2\rangle_{I_1\setminus J}
    > \langle z_1\rangle_{I_1\setminus J}.
 \ee
 where the first inequality holds because $z_2$ is decreasing to the left
of the origin and increasing to the right, and the second because
$z_2>z_1$ on $I_1\setminus J$.  But then
 \bea\label{eq:chain}
 \langle z_1\rangle_{I_1}
   &=&\frac\beta{\theta_1}r_B+\frac{\theta_1-\beta}{\theta_1}\langle
        z_1\rangle_{I_1\setminus J}\nonumber\\
   &=&r_B+\frac{\theta_1-\beta}{\theta_1}(\langle
        z_1\rangle_{I_1\setminus J}-r_B)\nonumber\\
   &<&r_B+\frac{\theta_1-\beta}{\theta_1}(\langle
        z_2\rangle_{I_2\setminus J}-r_B)\nonumber\\
   &<&r_B+\frac{\theta_2-\beta}{\theta_2}(\langle
        z_2\rangle_{I_2\setminus J}-r_B)\nonumber\\
  &=& \langle z_2\rangle_{I_2},
 \eea
  which contradicts \eqref{eq:third1}.
\end{proofof}

\section{Uniqueness of type 1 solutions}\label{sec:proofunique}

Our goal in this section is to prove

\begin{thm}\label{thm:unique} For any $\beta>0$ and any positive $r_A$,
$r_B$, and $r_C$ which satisfy $\sum_\alpha r_\alpha=1$ and are not all
equal to $1/3$ there exists at most one type~1 solution of \eqref{osc}
satisfying \eqref{eq:phases}--\eqref{eq:ralphay}. \end{thm}

 Throughout the section we will consider two solutions $y_{K_1},y_{K_2}$ of
\eqref{osc} with $0<K_2<K_1<1/27$; we write $\tau_i$ rather than $\tau_{K_i}$,
$i=1,2$, for the corresponding periods, and note that $\tau_1<\tau_2$ by
\crem{rem:trivial}(c).  We begin with a preliminary result.

\begin{lem}\label{lem:prelim} (a) For any $t_0\in\bR$ and $n\in\bZ$ the
curves $y_{K_1}(t+t_0)$ and $y_{K_2}(t)$ intersect exactly once in the
interval $(n\tau_2/2,(n+1)\tau_2/2)$.
 
 \smallskip\noindent
 (b) (i) If  $0\le t\le \tau_2/6$ then 
 \be\label{eq:lemi}
 y_{K_2}(2\tau_2/3-t)>y_{K_1}(2\tau_1/3-t),
 \ee
  and (ii) if in addition  $\tau_2\le3\tau_1$ then 
 \be\label{eq:lemii}
y_{K_2}(\tau_2-t)<y_{K_1}(\tau_1-t).
 \ee
\end{lem}

\begin{proof} (a)  Suppose that $n$ is even so that $y_{K_2}$ is increasing
  on the interval  $(n\tau_2/2,(n+1)\tau_2/2)$; the proof for $n$ odd is
similar.  For all $t$ and $t_0$ the zeros $a(K)$ and $b(K)$  of $U_K$
(see \eqref{eq:potential}) satisfy
 \be\label{eq:comp}
a(K_2)<a(K_1)\le y_{K_1}(t+t_0)\le b(K_1)<b(K_2),
 \ee
  and since $y_{K_2}(n\tau_2/2)=a(K_2)$ and $y_{K_2}((n+1)\tau_2/2)=b(K_2)$, the
existence of at least one intersection follows from the intermediate value
theorem.  On the other hand, if $y_{K_1}(t+t_0)=y_{K_2}(t)=y$ for some
$t\in(n\tau_2/2,(n+1)\tau_2/2)$ then from \eqref{eq:potential} and \eqref{osc},
$|y'_{K_1}(t+t_0)|=\sqrt{-2U_{K_1}(y)}<\sqrt{-2U_{K_2}(y)}=y'_{K_2}(t)$, and
this is inconsistent with the existence of two such intersection points.

 \smallskip\noindent
 (b) It follows from (a) that for any $t_1,t_2$, $y_{K_1}(t+t_1)$ and
$y_{K_2}(t+t_2)$ can intersect at most once on any interval of monotonicity
of $y_{K_2}(t+t_2$).  Thus it suffices to verify that \eqref{eq:lemi} and
\eqref{eq:lemii} hold for $t=0$ and $t=\tau_2/6$.  For \eqref{eq:lemi} this
follows from $y_{K_2}(\tau_1/2)=b(K_2)$ (see \eqref{eq:comp}) and
 \be
 y_{K_2}\Bigl(\frac{2\tau_2}3\Bigr)=\frac{1-a(K_2)}2>\frac{1-a(K_1)}2
    =y_{K_1}\Bigl(\frac{2\tau_1}3\Bigr).
 \ee
 For \eqref{eq:lemii} it follows from $y_{K_2}(\tau_2)=a(K_2)$ and
 \be
 y_{K_2}\Bigl(\frac{5\tau_2}6\Bigr)=\frac{1-b(K_2)}2<\frac{1-b(K_1)}2
    =y_{K_1}\Bigl(\frac{5\tau_1}6\Bigr)<y_{K_1}\Bigl(\tau_1-\frac{\tau_2}6\Bigr),
 \ee
 where at the last step we have used monotonicity of $y_{K_1}$ on
 $[\tau_1/2,\tau_1]$ and the condition $\tau_2\le3\tau_1$.  \end{proof}

In the remainder of this section we suppose
that, given appropriate phase shifts, the solutions $y_{K_1}$ and
$y_{K_2}$ considered above provide two type~1 solutions satisfying 
\eqref{eq:ralphay}--\eqref{eq:phases} for the given  $r_A$,
$r_B$, and $r_C$, and from this derive a contradiction.  It is
convenient to view these solutions in the same interval, which we take to
be $J=[-\beta,\beta]$.  This means that there are phase shifts $t_1$ and
$t_2$ such that if $z_i(t)=y_{K_i}(t+t_i)$ and
$w_i(t)=y_{K_i}(t+t_i+\tau_i/3)$ for $i=1,2$, then
 \be\label{eq:same}
 \langle z_1\rangle_J=\langle z_2\rangle_J=r_B \quad\hbox{and}\quad
 \langle w_1\rangle_J=\langle w_2\rangle_J=r_A.
 \ee
 By \crem{rem:averages}(c) we may again assume that
$r_A\ge r_C\ge r_B$ and that
 \be\label{eq:t1}
0\le t_i\le \frac{\tau_i}6, \qquad\hbox{$i=1,2$}.
 \ee
 The fact that these are type~1 solutions, and our assumption that
$K_1>K_2$, imply that
 \be\label{eq:t2}
   \beta<\frac{\tau_1}2<\frac{\tau_2}2.
 \ee

The key idea in the proof is the same as that for the simple case
considered in Section~\ref{sec:preunique}, but the general case presents
two additional difficulties.  First, without guidance from symmetry the
choice of the intervals $I_1$ and $I_2$ is more delicate, and in fact we
must consider two distinct cases in which they are chosen by different
prescriptions.  Second, one must in some cases apply the reasoning to $z_1$
and $z_2$, and in others to $w_1$ and $w_2$, and one must show that one or
the other is possible.  Sorting out these cases requires some detailed
analysis of the geometry of the curves.

\begin{proofof}{Theorem \ref{thm:unique}} 
 Equation \eqref{eq:same} implies that the curves $z_1(t)$ and
$z_2(t)$ must intersect at least once in the interior of the interval $J$,
and since the length $2\beta$ of $J$ is less than $\tau_2$,
\clem{lem:prelim}(a) implies that they cannot intersect more than three
times.  On the other hand, from \eqref{eq:int2} it follows that
$z_1(\beta)/z_2(\beta)=z_1(-\beta)/z_2(-\beta)$.  It is convenient then to
consider two possible alternatives:

 {\narrower
 \smallskip\noindent
 {\bf z1:} $z_2(\beta)>z_1(\beta)$, $z_2(-\beta)>z_1(-\beta)$, and the
 curves $z_1$ and $z_2$ intersect precisely twice within $J$;

\smallskip\noindent
 {\bf z2:} $z_2(\beta)\le z_1(\beta)$, $z_2(-\beta)\le z_1(-\beta)$, and
either both inequalities are strict and the curves intersect precisely
twice within $J$, or both are equalities and the curves have one
intersection within $J$.\par}

 \smallskip\noindent
A similar analysis leads to corresponding alternatives for $w_1$ and
$w_2$:

 {\narrower
 \smallskip\noindent
 {\bf w1:} $w_2(\beta)<w_1(\beta)$, $w_2(-\beta)<w_1(-\beta)$, and the
 curves $w_1$ and $w_2$ intersect precisely twice within $J$;

\smallskip\noindent
 {\bf w2:} $w_2(\beta)\ge w_1(\beta)$, $w_2(-\beta)\ge w_1(-\beta)$, and
either both inequalities are strict and the curves intersect precisely
twice within $J$, or both are equalities and the curves have one
intersection within $J$.\par}

 \smallskip\noindent
 We will first show that if {\bf z1} (respectively {\bf w1}) occurs then
the argument of Section~\ref{sec:preunique} may be applied to $z_1$ and
$z_2$ (respectively $w_1$ and $w_2$) to derive a contradiction.  Then
we show that either {\bf z1} or {\bf w1} or must occur.

 Now  observe that necessarily $\beta>t_2$.  For otherwise,
$0\le t_2-\beta<t_2+\beta\le \tau_2/3$, so that $z_2(t)=y_{K_2}(t+t_2)$ is
increasing on $J$ and by \clem{lem:prelim}(a) cannot intersect $z_1(t)$ more
than once there.  It is then helpful to further subdivide the situation
into two cases in which we can explicitly identify intervals on which $z_2$
is monotonic:

 {\narrower
 \smallskip\noindent
{\bf Case 1: $\beta+t_2>\tau_2/2$.} In this case we write
$J=J_1\cup J_2\cup J_3$, where $J_1=[-\beta,-t_2]$, $J_2=[-t_2,\tau_2/2-t_2]$,
and $J_3=[\tau_2/2-t_2,\beta]$.

 \smallskip\noindent
{\bf Case 2: $\beta+t_2\le \tau_2/2$.} Here we write $J=J_1\cup J_2$, where
$J_1=[-\beta,-t_2]$ and $J_2=[-t_2,\beta]$.\par}

 \smallskip\noindent
 Then $z_2$ is decreasing on $J_1$ and $J_3$ and increasing on $J_2$.

 Suppose now that {\bf z1} occurs.  Since $z_2(-\beta)>z_1(-\beta)$ and
$z_2(-t_2)=y_{K_2}(0)<z_1(-t_2)$ it follows from
\clem{lem:prelim}(a) that $z_1(t)$ and $z_2(t)$ intersect once on $J_1$.
We now consider separately the two cases introduced above.

 \smallskip\noindent
 {\bf Case~1:} Define $I_1=[\beta-\tau_1,\beta]$ and $I_2=[\beta-\tau_2,\beta]$,
so that $J\subset I_1\subset I_2$.  Since $I_i$ has length $\tau_i$, we again
have \eqref{eq:third1} and therefore \eqref{eq:ineq1}.  Moreover,
\eqref{eq:ineq2} follows from the monotonicity of $z_2$ on
$I_2\setminus J$ and then the fact that $z_2>z_1$ on
$I_1\setminus J$, which follows from this monotonicity,
\clem{lem:prelim}(a), and the fact that $z_2$ intersects $z_1$ on $J_1$.
But then we may deduce a contradiction between \eqref{eq:same} and
\eqref{eq:third1}, just as we did in  \eqref{eq:chain}.

  \smallskip\noindent
 {\bf Case~2:} The argument is similar.  Let
$I_2=[-t_2-\tau_2/2,\tau_2/2-t_2]=I_2^\ell\cup I_2^r$, where
$I_2^\ell=[-t_2-\tau_2/2,-t_2]$ and $I_2^r=[-t_2,\tau_2/2-t_2]$.  Let
$I_1=I_1^\ell\cup I_1^r$, where $I_1^\ell=[-a,-t_2]$ and $I_2^r=[-t_2,b]$,
be an interval of length $\tau_1$ satisfying $J\subset I_1\subset I_2$, with
$a$ and $b$ chosen so that
 \be\label{eq:condition}
\frac{|I_1^\ell\setminus J|}{|I_2^\ell\setminus J|}
  = \frac{|I_1^r\setminus J|}{|I_2^r\setminus J|}.
 \ee
 Now by arguing as in Case~1 we find that \eqref{eq:ineq2} holds with $I_2$
and $I_1$ replaced either by $I_2^\ell$ and $I_1^\ell$ or by $I_2^r$ and
$I_1^r$.  Averaging these two equations (the weights for this averaging,
 \be
\frac{|I_i^\ell\setminus J|}
   {|I_i^\ell\setminus J|+|I_i^r\setminus J|}
  \qquad\hbox{and}\qquad
 \frac{|I_i^r\setminus J|}
   {|I_i^\ell\setminus J|+|I_i^r\setminus J|}
 \ee
 are independent of $i$ by \eqref{eq:condition}) we again find that
\eqref{eq:ineq2} itself holds, and the argument proceeds to a contradiction
as in Case~1.

If {\bf w1} occurs then we may derive a contradiction similarly: we need
only replace $z_i$ by $w_i$ throughout and change the sign of some of the
inequalities.

It remains to show that either {\bf z1} or {\bf w1} must occur; to do so we
assume that {\bf z2} occurs and show that then {\bf w1} must also.  The
occurrence of {\bf z2} implies that we must be in Case~1, i.e., that
 \be\label{eq:case1}
\beta+t_2>\frac{\tau_2}2,
 \ee
  since if $z_2(-\beta)<z_1(-\beta)$ there can be no intersection of $z_1$
and $z_2$ on $J_1$ and hence the two intersections of these curves must
occur on $J_2$ and $J_3$, while if $z_2(-\beta)=z_1(-\beta)$ then there are
three intersections which must occur on $J_1$, $J_2$ and $J_3$.  We claim
further that necessarily
 \be\label{eq:keyineq}
  t_2-\frac{2\tau_2}3 > t_1-\frac{2\tau_1}3.
 \ee
 For if \eqref{eq:keyineq} does not hold then
\be
z_2(\beta)=y_{K_2}(\beta+t_2)
   \ge y_{K_2}\left(\beta+t_1+\frac{2(\tau_2-\tau_1)}3\right) 
  > y_{K_1}(\beta+t_1)=z_1(\beta).
\ee
 Here for the first inequality we have used \eqref{eq:t1}, \eqref{eq:t2},
\eqref{eq:case1}, and the falsity of \eqref{eq:keyineq}, which together
imply that
 \be\label{eq:goodineq}
   \frac{\tau_2}2 < \beta+t_2 \le \beta+t_1+\frac{2(\tau_2-\tau_1)}3 < \frac{2\tau_2}3,
 \ee
  and then the monotonicity of $y_{K_2}$ on $[\tau_2/2,2\tau_2/3]$.  For the
second inequality we use \clem{lem:prelim}(b.i), which is applicable by
\eqref{eq:goodineq}.  But now
 \bea
w_2(\beta)&=&y_{K_2}\left(\beta+t_2+\frac{\tau_2}3\right)\nonumber\\
   &<&  y_{K_1}\left(\beta+t_2+\tau_1-\frac{2\tau_2}3\right)\nonumber\\ 
  &<& y_{K_1}(\beta+t_1+\tau_1/3)=w_1(\beta),
 \eea
 which implies {\bf w1}.  Here the first inequality is from
\clem{lem:prelim}(b.ii), applicable because \eqref{eq:t1}, \eqref{eq:t2},
and \eqref{eq:case1} imply 
 \be\label{eq:final} 
0\le 2\tau_2/3-\beta-t_2\le \tau_2/6,
 \ee
  and \eqref{eq:case1} implies that $\beta>\tau_2/3$ and so $\tau_2<3\tau_1/2$, and
the second from \eqref{eq:keyineq} and the monotonicity of $y_{K_1}$ on
$[2\tau_1/3,\tau_1]$, which is applicable by \eqref{eq:final} and the
observation that \eqref{eq:case1} implies that $\beta>\tau_2/3>\tau_1/3$.  \end{proofof}

\section{Solutions of type $n$, $n\ge2$}\label{sec:typen}

It is easy to see that type $n$ solutions of \eqref{eq:ralphay} with
$n\ge2$ do exist for certain but not all values of $\beta$ and $\rhobar$.
The next result, however, shows that these solutions are not of physical
interest.

\begin{thm}\label{rearrange} No type~$n$ solution of \eqref{eq:ralphay}
with $n\ge2$ minimizes the free energy.  \end{thm}

We begin by rewriting the free energy in terms of the variable
$t=2\beta x$.  Consider then triples $z(t)=(z_A(t),z_B(t),z_C(t))$ of
functions, defined in some interval $[c,d]$ of length $2\beta$ and
satisfying $0<z_\alpha(t)<1$ and $\sum_\alpha z_\alpha(t)=1$.  The free
energy $\F(\{z(t)\})$ of $z$ is (see \eqref{eq:F})
 \be
\F(\{z(t)\})=(4\beta)^{-1}(\E(\{z(t)\})-\S(\{z(t)\})),
 \ee
where
 \be
\E(\{z\})=\int_c^d dt\int_t^d ds\,[z_A(t)z_C(s)+z_B(t)z_A(s)+z_C(t)z_B(s)]
 \ee
is the energy and
 \be
\S(\{z\})=-2\int_c^d dt\,[z_A\log z_A+z_B\log z_B+z_C\log z_C]
 \ee
 the entropy.  We will use  a {\it rearrangement} procedure for
these triples.

Let $c=t_0<t_1<\cdots<t_m=d$ be a partition of $[c,d]$ and for
$k=1,\ldots,m$ let $I_k=(t_{k-1},t_k]$.  Given a permutation $\sigma$ of
$\{1,2,\ldots,m\}$ we may rearrange the intervals $I_k$ in $[c,d]$,
together with the restrictions of $z$ to each interval, into the order
$I_{\sigma(1)},\ldots,I_{\sigma(m)}$, thus defining in the obvious way a
new function $w$ on $[c,d]$, the {\it rearrangement} of $z$.  A formal
definition of $w$ (which we will not use in the sequel) may be given as
follows: The permuted intervals arise from the partition
$c=s_0<s_1<\cdots<s_m=d$, with
$s_j=\sum_{\{i\mid\sigma(i)\le j\}}(t_i-t_{i-1})$, and this permutation is
implemented by the map $\psi:[c,d]\to[c,d]$ defined by
 \be
\psi(c)=c,\qquad \psi(x) = x-t_{k-1}+s_{k-1} \quad\hbox{if $x\in I_k$}. 
 \ee
 The rearrangement of $z$ is then $w=z\circ\psi^{-1}$, i.e.,
$w_\alpha(\psi(t))=z_\alpha(t)$.

The entropy of $z$ and $w$ are the same.  The energy of $z$ is
 \be
\E(\{z(t)\})=\sum_{i=1}^m\E_i+\sum_{1\le i<j\le m}\E_{ij}
 \ee
 where 
 \be
 \E_i=\int_{t\in I_i} dt\int_{s\in I_i, s>t}ds\,
      [z_A(t)z_C(s)+z_B(t)z_A(s)+z_C(t)z_B(s)],\\
 \ee
 and for $i<j$, with $\zbar_{i\alpha}=\int_{I_i}z_\alpha(t) \,dt$, 
 \begin{eqnarray} \label{eij}
\E_{ij}&=&\int_{t\in I_i}dt\int_{s\in I_j} ds\,
  [z_A(t)z_C(s)+z_B(t)z_A(s)+z_C(t)z_B(s)]\\\nonumber
  &=&\zbar_{i,A}\zbar_{j,C}+\zbar_{i,B}\zbar_{j,A}+\zbar_{i,C}\zbar_{j,B}.
 \end{eqnarray}
 It is natural to think of \eqref{eij} as expressing $\E(\{z(t)\})$ as a
sum of ``self energies'' of the restrictions of $z$ to the intervals $I_i$
and ``interaction energies'' between the portions of $z$ in different
intervals.  The energy $\E(\{w(t)\})$ is obtained similarly; the self
energy contribution will be the same and the interaction energies will
differ only for interval pairs $I_i,I_j$ whose order is interchanged by the
rearrangement; thus
 \bea
\F(\{w(t)\})-\F(\{z(t)\})
&=&\E(\{w(t)\})-\E(\{z(t)\})\nonumber\\
   &=&\sum_{\{i,j\mid i<j,\;\sigma(j)<\sigma(i)\}}\Delta\E_{i,j},
\label{eq:rearr}
  \eea
  where
 \be \label{delta}
\Delta\E_{i,j}=[\zbar_{i,C}\zbar_{j,A}+\zbar_{i,A}\zbar_{j,B}
    +\zbar_{i,B}\zbar_{j,C}-\zbar_{i,A}\zbar_{j,C}
    -\zbar_{i,B}\zbar_{j,A}-\zbar_{i,C}\zbar_{j,B} ].
 \ee

\begin{figure}[ht!]  
\begin{center}
\includegraphics[width=10cm, height=6cm]{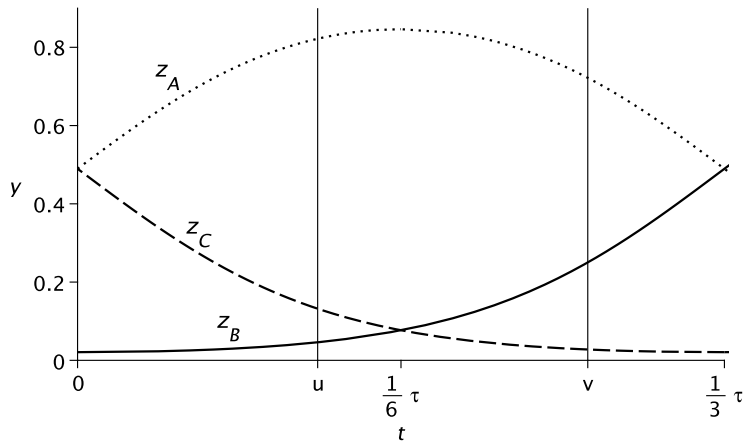}
\caption{Plots of $z_A$, $z_B$, and $z_C$ in the interval $[0,\tau_K/3]$ for
  $K=0.005$, with typical values of $u$ and $v$.} 
\label{fig:typen} 
\end{center}
\end{figure}

 \begin{proofof}{\cthm{rearrange}} Suppose that the triple $z(t)$ as above
gives a solution of \eqref{eq:ralphay} on the interval $[c,d]$;
specifically, this means that for some $K$ and some phase shift $t_B$,
$c=t_B-\beta$, $d=t_B+\beta$, and $z_A(t)=y_K(t+\tau_K/3)$, $z_B(t)=y_K(t)$,
and $z_C(t)=y_K(t-\tau_K/3)$.  Suppose also that the solution is of type $n$
for some $n\ge2$, which implies in particular that $\tau=\tau_K<2\beta$.  We will
show that for some rearrangement $w$ of $z$ the change in free energy
\eqref{eq:rearr} is negative.  Since clearly
$\langle w_\alpha\rangle_{[c,d]}=\langle z_\alpha\rangle_{[c,d]}$ for all
$\alpha$, $w$ will satisfy the same constraints as $z$; thus  $z$ cannot
be a minimizer of  the free energy under those constraints.

The desired rearrangement will be defined in terms of a partition
$t_0,\ldots,t_m$ with $t_0=c$, $t_k=(k_0+k)\tau/3$ for $k=1,\ldots,m-1$, and
$t_m=d$; here $k_0$ is the largest integer such that $k_0\tau/3\le c$ and $m$
is the smallest integer such that $(k_0+m)\tau/3\ge d$.  Note that here $I_1$
and $I_m$ have length at most $\tau/3$ and $I_j$, $j=2,\ldots,m-1$, has length
exactly $\tau/3$.  It is easy to see that $(m-2)\tau/3<2\beta\le m\tau/3$, so that
from \eqref{eq:typedef}, $3n-2\le m\le3n+1$ and in particular $m\ge4$.

Now for each $j$, $j=1,\ldots,m$, there will be an $\alpha\in\{A,B,C\}$
such that $z_\alpha(t)\ge z_{\alpha\pm1}(t)$ for all $t\in I_j$ (see
Figure~\ref{fig:solns}); we will then say that $I_j$ is a {\it full
$\alpha$-interval} if it has length $\tau/3$ and a {\it partial
$\alpha$-interval} if it has length less than $\tau/3$ (which is possible only
if $j=1$ or $m$).  The types of the intervals are in cyclic order (again
see Figure~\ref{fig:solns}): if $I_j$ is an $\alpha$-interval then
$I_{j+1}$ is an $(\alpha+1)$-interval.  Recalling that $m\ge4$ we focus on
four consecutive intervals $I_i,\ldots,I_{i+3}$.  Without loss of
generality we may assume that $I_i$ is an $A$-interval, and we will then
write $I_i=A_l$, $I_{i+1}=B$, $I_{i+2}=C$, and $I_{i+3}=A_r$, i.e., the
initial configuration is $A_lBCA_r$.  Consider then the permutation of
these intervals in which we first exchange $B$ and $C$, then interchange
$A_l$ with $C$ (they are now adjacent) and $A_r$ with $B$, and finally
exchange $A_l$ and $A_r$, leading to $C,A_r,A_l,B$.  We will show that this
rearrangement lowers the energy, completing the proof of the theorem.

The calculation is simplest when all four of the intervals are full, so we
consider that special case first.  For a full $\alpha$-interval $I_j$ we
have $\zbar_{j,\alpha}=\zeta$ and $\zbar_{j,\alpha\pm1}=\hat\zeta$, where
 \be\label{full}
\zeta=\int_0^{\tau/3}z_A(t)\,dt,\qquad\hbox{and}\qquad
\hat\zeta=\int_0^{\tau/3}z_B(t)\,dt=\int_0^{\tau/3}z_C(t)\,dt,
 \ee
 and so $\zeta>\hat\zeta$.  We may then compute from \eqref{delta} and
\eqref{full} the energy difference $\Delta\E_{ij}$ arising from reversing
the order of a full $\alpha_i$-interval $I_i$ and a full
$\alpha_j$-interval $I_j$, where $i<j$: if $\alpha_j=\alpha_i$ then
$\Delta\E_{ij}=0$, while if $\alpha_j=\alpha_i\pm1$ then
$\Delta\E_{ij}=\pm e$, where $e=(\zeta-\hat\zeta)^2>0$.  Rearranging the
intervals as above thus first increases the energy by $e$, then decreases
it by $2e$; the final exchange of $A_l$ and $A_r$ leaves the energy
unchanged.  The net change of energy is $-e$.

We now consider the general case in which either $A_l$ or $A_r$ or both may
be partial intervals.  After a translation, if necessary, of $[c,d]$ by a
multiple of $\tau$ we may assume that $A_l=(u,\tau/3]$, $B=(\tau/3,2\tau/3]$,
$C=(2\tau/3,\tau]$, and $A_r=(\tau,\tau+v]$ for some $u,v$ with $0\le u<v\le \tau/3$; the
condition $u<v$ comes from the fact that if $A_r$ is full then $u=0$ and if
$A_l$ is full then $v=\tau/3$, while if both are partial then
$v+\tau-u=d-c=2\beta>\tau$.  The energy change from the rearrangement is the sum,
with $j=i+3$, of
\begin{itemize}
\item$(\zeta-\hat\zeta)^2$ from the exchange of $C$ and $B$,
\item$-(\zeta-\hat\zeta)(\zbar_{i,A}-\zbar_{i,B})$ from the exchange of
$A_r$ and $C$, 
\item$-(\zeta-\hat\zeta)(\zbar_{j,A}-\zbar_{j,C})$
from the exchange of $B$ and $A_r$, and
\item$\zbar_{i,C}\zbar_{j,A}+\zbar_{i,A}\zbar_{j,B}
+\zbar_{i,B}\zbar_{j,C}-\zbar_{i,A}\zbar_{j,C}
-\zbar_{i,B}\zbar_{j,A}-\zbar_{i,C}\zbar_{j,B}$ from the exchange of $A_l$
and $A_r$.  
\end{itemize}
This energy may conveniently be expressed in terms of
 \be
 \eta_\alpha^{(1)}=\int_0^uz_\alpha(t)\,dt, \quad
 \eta_\alpha^{(2)}=\int_u^vz_\alpha(t)\,dt, \quad\hbox{and}\quad
 \eta_\alpha^{(3)}=\int_v^{\tau/3} z_\alpha(t)\,dt,
 \ee
 using
 \be
 z_{i,\alpha}=\eta_\alpha^{(2)}+\eta_\alpha^{(3)},\quad
z_{j,\alpha}=\eta_\alpha^{(1)}+\eta_\alpha^{(2)},\quad
\zeta=\eta_A^{(1)}+\eta_A^{(2)}+\eta_A^{(3)},
 \ee
and
 \be\label{ident}
  \hat\zeta=\eta_B^{(1)}+\eta_B^{(2)}+\eta_B^{(3)}
  =\eta_C^{(1)}+\eta_C^{(2)}+\eta_C^{(3)}
 \ee
 (see Figure~\ref{fig:typen}).  Finally, if we introduce the strictly
positive quantities $\theta_B^{(k)}=\eta_A^{(k)}-\eta_B^{(k)}$ and
$\theta_C^{(k)}=\eta_A^{(k)}-\eta_C^{(k)}$, then the total energy change
may be written, using the identity displayed in \eqref{ident}, in the
manifestly negative form 
 \be
 -\theta_B^{(2)}\theta_C^{(3)}
 -\theta_C^{(2)}\theta_B^{(1)}
 -\theta_B^{(2)}\theta_C^{(2)}.
 \ee
\end{proofof}

\begin{rem}\label{rem:typen} Energy considerations for full intervals
correspond closely to those of single particles: cyclic order $ABC$ is
energetically favored, the energy cost of a reversed pair $BA$, $CB$, or
$AC$ is independent of the species involved, etc.  The conclusion above
that $CAAB$ is favored over $ABCA$ parallels the conclusion of
Section~\ref{sec:groundstate} that the ground state of a particle system
with a majority of $A$'s has a block containing all the $A's$ between
blocks of the $C$'s and the $B$'s.  For partial intervals, energy
considerations are more subtle; the final step in the rearrangement above,
the interchange of $A_l$ and $A_r$, is irrelevant when these intervals are
full but is needed when they are partial because $A_r$ is richer in species
$C$ than in $B$ and the reverse is true for $A_l$.  \end{rem}

\section{Phase diagram of the ABC model\label{sec:phased}}

\begin{figure}
\centerline{\includegraphics[width=10cm]{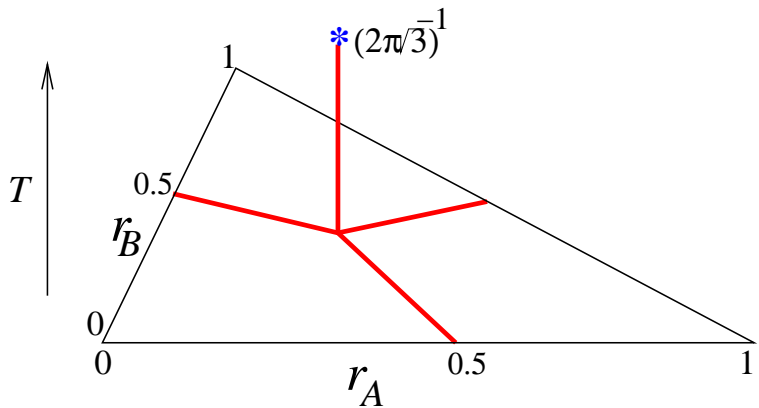} }
 \caption{The $(T,r_A, r_B)$  phase diagram: The temperature is added as a
vertical axis to the right triangle. The $T=0$ phase transition
lines do not extend to $T\ne 0$, except at the equal density point
$r_\al=1/3$ where it extends as a first-order line at low $T$ and
ends at the critical point $\beta = 2\pi\sqrt 3$.}
\label{fig:3Dphase}
\end{figure}

Our results about the minimizers of $\F$ for specified $\beta$, $r_A$, and
$r_B$ provide the following picture of the canonical phase diagram of the
model (see Fig. \ref{fig:3Dphase}). The three first order lines in the
$T=0$ plane, discussed in Section~\ref{sec:model}, do not extend to finite
temperatures.  There is a first order line for $r_A=r_B=r_C=1/3$ which
starts at $T=0$ and terminates at the critical temperature
$T_c=(2\pi\sqrt3)^{-1}$, i.e.,~as one crosses this line by varying the
overall densities $r_{\alpha}$, the density profiles change discontinuously
for $T<T_c$. On the line of equal densities the transition at $T_c$ is
second order; the divergence of the variance in local density fluctuations
as one approaches $T_c$ from above on this line was investigated by
Bodineau et al.~\cite{BDLvW} via a study of the $1/N$ dependence of the
correlation function.

\begin{figure}
\includegraphics*[width=10cm]{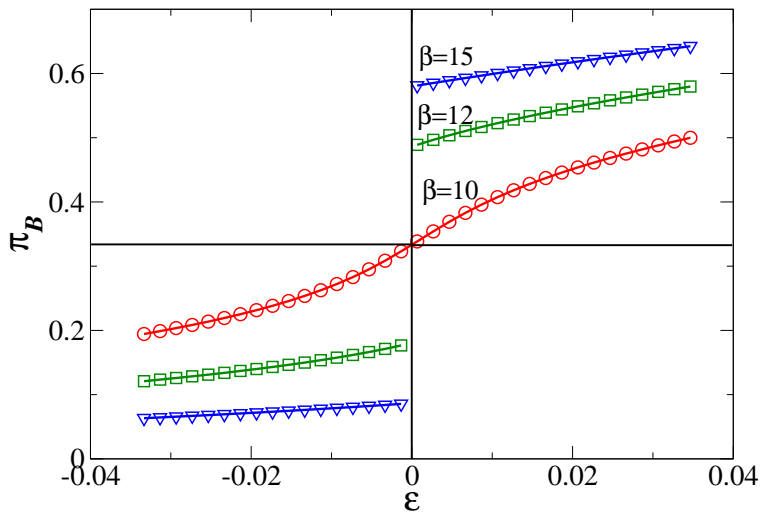}
\caption{ The order parameter $\pi_B$  defined in (\ref{eq:piB})
is shown as function of $\epsilon \equiv (r_B-1/3)/2$ for different
temperatures $\beta=10,12$ and $15$. Other parameters are, $r_A= r_C
= (1-r_B)/2$.  Data were obtained via a mean-field dynamical solution in a
system of 200 sites.} \label{fig:piB}
\end{figure}

To see how the density profiles change as a function of the average
densities near the $r_{\alpha}=1/3$ line we take $r_A=r_C$ and vary $r_B$.
The density profiles change continuously at high temperatures
$\beta<\beta_c$ as the profiles become uniform at $r_\al=1/3$. However,
when $\beta>\beta_c$, the $B$ particles prefer to be in the middle section
of the interval for $r_B>1/3$ and to be  symmetrically arranged at both
ends for $r_B<1/3$, and this  change of profile is
discontinuous. To see this discontinuity we define an order parameter,
 \be \pi_B = 2 \int_{1/4}^{3/4} \rB(x) dx, \label{eq:piB} \ee 
and calculate it in the steady state. Figure \ref{fig:piB}, shows the order
parameter $\pi_B$ as a function of $r_B$ for $\beta=10,12$ and $15$.
Clearly, for low temperatures, $\beta>\beta_c$, $\pi_B$ shows a
discontinuity at density $r_B=1/3$.

\subsection{Perturbation expansion}
\label{perturbation}
In this section we apply a perturbation expansion to calculate the
density profiles for average densities which deviate slightly from
$1/3$. In particular we carry out a perturbation expansion of the
profiles for average densities $r_A = 1/3+\eps+\eta$ , $r_B= 1/3
-2\eps$ and $r_C=1/3 +\eps-\eta$ with small parameters $\epsilon$
and $\eta$~. For $\epsilon=\eta=0$ the density profiles are
homogeneous (namely, $\rho_\alpha(x)=1/3$) at high temperatures, and
they become non-homogeneous at $\beta<\beta_c= 2\pi \sqrt 3$~. We
now expand the density profiles around the homogeneous densities:
\be
\rho_A(x) = \frac 13 + a(x), ~~ \rho_B(x) = \frac 13 + b(x), ~~
{\rm and}~ \rho_C(x) = \frac 13 + c(x)
\ee
with small deviations $a(x), b(x)$ and $c(x)$ which satisfy $a(x) +
b(x) + c(x) =0$. As a result of this constraint, one is left with
two independent functions which we choose to be $b(x)$ and
$s(x)\equiv a(x) -c(x)$. The average densities are then  fixed by,
 \be \int_{-1/2}^{1/2} b(x) dx =-2\eps ~~{\rm and}
~~\int_{-1/2}^{1/2} s(x) dx= 2\eta \label {eq:beps}, \ee
where, for convenience, the interval is taken to be $-1/2 \le x \le
1/2$. In terms of $b(x)$ and $s(x)$ the equations for the minimizers
(\ref{eq:dABC}) reduce to
\bea
\frac {db}{dx} &=& \frac \beta 3  s + \beta b s \cr
\frac {ds}{dx} &=& -\beta b + \frac{3\beta}{2} b^2 - \frac \beta 2 s^2
\label{eq:bd}
\eea

We proceed by expanding $b(x)$  and $s(x)$  in terms of two small
parameters $u$ and $v$, which, at the end of the calculation, will
be determined by $\epsilon$ and $\eta$~. To be explicit, let us
write
\be b(x) =\sum_{k=1}^{\infty} b_k(x),~{\rm and}~~ s(x) =
\sum_{k=1}^{\infty} s_k(x), \ee
where $b_k(x)$ and $s_k(x)$ are functions of $x$ which are of order
$k$ in the small parameters $u$ and $v$. To first order in $u$ and
$v$ (\ref{eq:bd}) yields the following equations for $b_1(x)$
and $s_1(x)$
\be \frac {db_1}{dx} = \frac \beta 3 s_1  ~~~{ \rm and }~~~ \frac
{ds_1}{dx} = -\beta b_1 ~.\nonumber \ee
Solving these equations one finds
 \bea b_1(x) &=& u\cos (\alpha x)+v\sin(\alpha x) \cr s_1(x) &=& - u\sqrt 3 \sin(
 \alpha x)+v\sqrt{3} \cos(\alpha x)
\eea
with $\alpha = \beta/\sqrt{3}$~. Expanding $b(x)$ and $s(x)$ to
second order in $u$ and $v$ the following equations are obtained for
$b_2(x)$ and $s_2(x)$,
\bea
 \frac{d b_2 }{dx} &=& \frac {1}{\sqrt 3}\alpha s_2 -\frac{ 3\alpha}{2}
(u^2-v^2)\sin(2\alpha x)+ 3\alpha uv \cos(2\alpha x) \cr
 \frac{d s_2}{dx} &=& -\sqrt{3} \alpha b_2 +  \frac{3\sqrt{3}\alpha}{2}(u^2-v^2)  \cos(2\al
 x)+3\sqrt{3} \alpha uv \sin(2\alpha x)~. \eea
These equations can be solved to yield
\bea
 b_2(x) &=& \frac 12 (u^2-v^2)\cos (2\al x)+uv\sin(2\al x) \cr
 s_2(x) &=& \frac{\sqrt 3}{2}(u^2-v^2) \sin(2\al x)-\sqrt{3}uv\cos(2\al x).
\eea
The equations for the third order terms, $b_3(x)$ and $s_3(x)$~, are
then
\bea \label{eq:b3s3} \frac{db_3}{dx} &=& \frac {1}{\sqrt 3}\al s_3+
\frac{3}{2} \alpha uw^2\sin(\alpha x)-\frac{3}{2}\al vw^2\cos(\al x)
\cr \frac{d s_3}{dx} &=& -\sqrt{3}\al b_3 + \frac{3\sqrt{3}}{2}
\alpha uw^2\cos(\al x)+\frac{3\sqrt{3}}{2}\al vw^2\sin(\al x)~, \eea
where $w^2=u^2+v^2$. To proceed with the analysis one eliminates
$s_3(x)$ from the equation for $b_3(x)$ to obtain the following
second order differential equation for $b_3(x)$
\begin{equation}
\label{eq:b3} \frac{d^2b_3}{dx^2} + \alpha^2b_3 = 3\al^2
uw^2\cos(\alpha x)+3\al^2 vw^2\sin(\al x)~.
\end{equation}
This equation may be solved to yield

\be
 b_3(x) = \frac 32 \alpha uw^2 x \sin(\alpha x)-\frac 32 \al vw^2x\cos(\al x)~.
\ee
Inserting this solution in (\ref{eq:b3s3}) results in the
following solution for $s_3$

\begin{equation}
s_3(x)= \frac{3\sqrt{3}}{2} \alpha uw^2 x \cos(\alpha
x)+\frac{3\sqrt 3}{2}\al vw^2\sin(\al x)~.
\end{equation}
In summary, to third order in $u$ the density profiles are given by
\bea   b(x) &=& u \cos(\al x) +v\sin(\al x) + \frac{1}{2} (u^2-v^2)
\cos(2\al x) +uv\sin(2\al x) \cr &&\hskip20pt +\; \frac 32 \alpha uw^2 x
\sin(\alpha x)- \frac 32 \alpha vw^2 x \cos(\alpha x) ,\cr s(x) &=&
-u \sqrt 3 \sin(\al x) +v\sqrt 3 \cos(\alpha x) + \frac{\sqrt3}{2}
(u^2-v^2) \sin(2\al x) \cr &&\hskip20pt-\;
 \sqrt{3} uv\cos(2\al x) +\frac{3\sqrt{3}}{2} \alpha uw^2 x \cos(\alpha
 x)\cr &&\hskip20pt+\;
\frac{3\sqrt{3}}{2} \alpha vw^2 x \sin(\alpha x). \label{eq:profiles}
\eea
To third order in $u$ these equations may be reexpressed as
\bea   b(x) &=& u \cos(\gamma x)  + v\sin(\gamma x) + \frac{1}{2}
(u^2-v^2) \cos(2\gamma x)\cr &+& uv\sin(2\gamma x) + O(u^4,v^4)~
,\cr s(x) &=& -u \sqrt 3 \sin(\gamma x )+v\sqrt{3}\cos(\gamma x) +
\frac{\sqrt3}{2} (u^2-v^2) \sin(2\gamma x) \cr &-& \sqrt 3
uv\cos(2\gamma x)+O(u^4,v^4)~, \label{eq:profiles1} \eea
with
\be \gamma=\alpha -\frac{3}{2}\alpha w^2. \label{eq:gamma}\ee
Note that $s(x)=b(x+2\pi/3\gamma)-b(x-2\pi/3\gamma)$, which is consistent
with \eqref{solns}--\eqref{eq:phases}.

The expansion parameters $u$ and $v$ may be expressed in terms of
$\epsilon$ and $\eta$ using (\ref{eq:beps}). This leads to
\bea
 -2\eps &=& \left[\frac{2}{\gamma}
\sin(\gamma/2) \right] u + \left[\frac{1}{2\gamma} \sin{\gamma}\right]
(u^2-v^2) \cr \cr  2\eta &=& \left[\frac{2\sqrt
3}{\gamma} \sin(\gamma/2) \right] v - \left[\frac{\sqrt 3}{\gamma}
\sin{\gamma}\right]uv 
~.\label{eq:orderparameter1} \eea
 For $\epsilon=\eta=0$ a nonzero solution of \eqref{eq:orderparameter1}
exists only if $\sin\gamma/2=0$, and thus by \eqref{eq:gamma} only if
$\alpha>2\pi$, i.e., $\beta>\beta_c$. For $\beta\le\beta_c$, then, all
profiles are homogeneous.  For $\beta>\beta_c$ there exist type $n$
solutions with $\gamma=2n\pi$ and $w$ small whenever $\beta$ is just above
$n\beta_c$.  Their amplitude $w=(u^2+v^2)^{1/2}$ is determined by
$\beta$ via \eqref{eq:gamma} but  $u$ and $v$ are otherwise
undetermined, corresponding to the translation invariance of the set of
solutions.

For non-vanishing $\epsilon$ or $\eta$ it is convenient to expand
\eqref{eq:orderparameter1} to third order:
\bea
 -2\eps &=& \left[\frac{2}{\al}
\sin(\al/2) \right] u + \left[\frac{1}{2\al} \sin{\al}\right]
(u^2-v^2) \cr \cr &+& \frac 32 \left[-\cos(\al/2) +  \frac{2}{\al}
\sin(\al /2) \right] uw^2 \cr \cr 
  2\eta &=& \left[\frac{2\sqrt
3}{\al} \sin(\al/2) \right] v - \left[\frac{\sqrt 3}{\al}
\sin{\al}\right]uv \cr \cr &+& \frac {3\sqrt 3}{2}
\left[-\cos(\al/2) +  \frac{2}{\al} \sin(\al /2) \right] vw^2
~.\label{eq:orderparameter} \eea
Then (\ref{eq:orderparameter}) has one or more non-vanishing solutions for
$u$ or $v$ at any temperature. The equilibrium (minimizing) solution is the
one which is of type $1$.  The minimizing density profiles, as has been
shown, are non-homogeneous both above and below $\beta_c$; they vary
continuously with the temperature $\beta$, and no phase transition takes
place.

The small $u$ and $v$ expansion may be used to verify the first order
nature of the transition at $\epsilon=\eta=0$ just below the critical
point. To demonstrate this point we take, for simplicity, $\eta=0$ and
consider small $\epsilon$ and $\alpha =2\pi +\Delta \alpha$ with
$\Delta \alpha >0$. In this case $v=0$, the amplitude $u$ satisfies
\begin{equation}
2\epsilon=\frac{1}{2\pi}(\Delta \alpha-3\pi u^2)u ~,
\label{u-epsilon}
\end{equation}
and the density profile is given by
\begin{equation}
b(x)=u \cos(2\pi + \Delta \alpha - 3 \pi u^2)x ~.
\end{equation}
At $\epsilon=0$, equation (\ref{u-epsilon}) has three solutions,
\begin{equation}
u=0 \qquad , \qquad  \pm \sqrt{\frac{\Delta \alpha}{3\pi}}~.
\end{equation}
Since the stable solution satisfies $\Delta \alpha -3 \pi u^2<0$, it
follows from (\ref{u-epsilon}) that for the stable solution one has
$\epsilon/u<0$. Thus for $\epsilon >0$ the negative $u$ solution is
stable and $\lim_{\epsilon\searrow0}u=-\sqrt{\Delta \alpha/3\pi}$, 
while for $\epsilon <0$ the positive $u$ solution is stable
and $\lim_{\epsilon\nearrow0}u=\sqrt{\Delta \alpha/3\pi}$.
The density profile is therefore discontinuous at $\epsilon=0$; see 
Figure~\ref{fig:piB}.

\section{Convexity of the free energy at high   temperature}\label{sec:convex}
\renewcommand{\rA}{n_A}\renewcommand{\rB}{n_B}\renewcommand{\rC}{n_C}

It follows from \cthm{thm:main} that if $\beta<\beta_c=(2\pi\sqrt{3})$ then
for any positive $r_A$, $r_B$ and $r_C$ with $\sum_\alpha r_\alpha=1$ there
is a unique solution of the ELE \eqref{eq:dABC} with the constraints
\eqref{eq:ralpha}; that is, a unique stationary point of the free energy
$\F$ under those constraints.  One would then like to know also the nature
of the fluctuations about this equilibrium profile, and in particular to
show that these fluctuations are Gaussian with a finite covariance.  To
establish this it is necessary to show that the free energy has a
positive definite second variation at the minimizer in the space of density
profiles satisfying the above constraints, which implies that it is a
strictly convex functional in a neighborhood of the minimizer.  We believe
that this is indeed true generally, with the obvious exception of the line
segment $\beta\ge\beta_c$, $r_A=r_B=r_C=1/3$.  In this section we show
a global convexity property in a restricted temperature range: for
$\beta < 4\pi/3=(4/3\sqrt3)\beta_c$, $\F$ (with the mean density of species
$\alpha$ constrained to be $r_\alpha$) is globally strictly convex.

To establish the convexity we compute the second variation. Consider the
free energy functional $\F(\{n(x)\})$ on the convex set of nonnegative
densities satisfying
 \be
 \rA(x) + \rB(x) + \rC(x) = 1 \qquad{\rm and}\qquad \int_0^1
n_\alpha(x){\rm d}x = r_\alpha.
 \ee
 To make the variations, let $\pA$ and $\pB$ be bounded continuous
functions with
 \be
 \int_0^1\pA(x){\rm d}x = \int_0^1\pB(x){\rm d}x = 0\ .
 \ee
Then for small values of $t$,
 \be
(\rA,\rB,\rC) \to (\rA+t\pA,\rB+t\pB,\rC -t[\pA+\pB])
 \ee
  is an admissible
variation, and all admissible variations are of this form.

The order $t^2$ contribution from the entropy is
\begin{multline}\label{eq:entbd}
   \frac{1}{2}\int_0^1 {\rm d}x \left[\frac{1}{\rA(x)} \pA^2(x) 
     +  \frac{1}{\rB(x)}\pB^2(x) 
     +  \frac{1}{\rC(x)}(\pA(x)+\pB(x))^2\right] \\
  \ge \int_0^1 {\rm d}x\left[\pA^2(x) + \pB^2(x)\right]\ ,
\end{multline}
where we have used the fact that for any positive $a$, $b$, and $c$ 
with $a+b+c=1$ the matrix
 \be
\begin{bmatrix}
   a^{-1}+c^{-1}-2&c^{-1}\\ c^{-1}&b^{-1}+c^{-1}-2
  \end{bmatrix}
 \ee
 is positive semidefinite, since its diagonal entries are nonnegative and
its determinant is $(abc)^{-1}[(a-b)^2(1-c)+c(1-c/2)^2+3c^3/4]$.  Because
the energy is quadratic, the order $t^2$ contribution from the energy is
independent of $(\rA,\rB,\rC)$; it is
 \begin{multline}\label{eq:enord2}
\beta \int_0^1{\rm d}x \int_0^1{\rm d}y \,
   \Theta(y-x)\bigl[-\pA(x)(\pA(y)+\pB(y)) \\
 +\pB(x)\pA(y) -(\pA(x)+\pB(x))\pB(y)\bigr]\ .
 \end{multline}
 However, since for functions $f$ and $g$ with $\int_0^1f(x){\rm d}x =
 \int_0^1g(x){\rm d}x =0$,
 \be\label{ident1}
\int_0^1{\rm d}x \int_0^1{\rm d}y \,\Theta(y-x)f(x)g(y) = - 
\int_0^1{\rm d}x \int_0^1{\rm d}y \,\Theta(y-x)g(x)f(y)\ ,
 \ee
 \eqref{eq:enord2} reduces to
 \begin{equation}\label{energy}
 3\beta \int_0^1{\rm d}x \int_0^1{\rm d}y \, \Theta(y-x)\pB(x)\pA(y)\ ,
 \end{equation}

Now let ${\cal H}$ be the Hilbert space of square integrable functions $f$
on $[0,1]$ with $\int_0^1f(x){\rm d}x = 0$, and   define on operator $K$ on
${\cal H}$ 
 \be
  Kf(x) = P\left(\int_x^1 f(y){\rm d}y\right)\ ,
 \ee
 where $P$ is the orthogonal projection onto ${\cal H}$ in $L^2([0,1])$.
Then it is easy to see that
 \begin{equation}\label{ee}
   \int_0^1{\rm d}x \int_0^1{\rm d}y \Theta(y-x)\pB(x)\pA(y) 
        = \langle \pB,K\pA\rangle_{{\cal H}}\ .
 \end{equation}
Combining the entropy bound \eqref{eq:entbd} with (\ref{energy}) and
(\ref{ee}), we see that the second variation is bounded below by
\begin{equation}\label{total}
   \int_0^1 {\rm d}x (\pA^2(x) + \pB^2(x)) 
   + 3\beta\langle \pB,K\pA\rangle_{\cal H}\ .
\end{equation}
 We need to show that this quadratic form is nonnegative definite for
$\beta < 4\pi/3$, which we shall do with a spectral calculation.

To do the calculation, define
 \be
\varphi_n(x) = \sqrt{2}\sin(n 2\pi x)\qquad{\rm and} \qquad 
   \psi_n(x) = \sqrt{2}\cos(n 2\pi x)\ .
 \ee
 Then one has
 \begin{equation}\label{gf}K \varphi_n(x) = \frac{1}{n2\pi} \psi_n(x)
   \qquad{\rm and} \qquad K \psi_n(x) = -\frac{1}{n2\pi}\varphi_n(x)\ .
\end{equation}
 If we now write
 \be
\pA = \sum_{n=1}^\infty (a_n\varphi_n + b_n\psi_n)\qquad{\rm and}\qquad 
\pB = \sum_{n=1}^\infty (c_n\varphi_n + d_n\psi_n)\ ,
 \ee
 then (\ref{total}) becomes
 \be
 \sum_{n=1}^\infty \left[(a_n^2+ b_n^2 + c_n^2+d_n^2) +
     \frac{3\beta}{n2\pi}(-c_n b_n + d_n a_n)\right]\ .
 \ee
 This is nonnegative as long as $\beta \le 4\pi/3$, which proves
the strict positivity of the second variation, and hence the convexity of
$\F$ for such $\beta$

We also remark that if in the equal density case
$r_A = r_B = r_C = \frac{1}{3}$ one considers quadratic variations around
the constant profile then one can improve the lower bound \eqref{eq:entbd}
on the second variation of the entropy to
 \be
3\int_0^1(\pA^2+\pB^2+\pA\pB)\,dx.
 \ee
   Some modification of the remainder of the argument then shows that the
constant profile is indeed a local minimum all the way to $\beta_c$.

\section{Existence of minimizers}\label{sec:existence}

In this section we prove \cthm{thm:exist}, that is, we show that there
exist profiles $\rho(x)$ which minimize the free energy functional
$\F(\{n(x)\})$ of \eqref{eq:F} and that these satisfy the corresponding
Euler Lagrange equations.  The latter cannot be taken for granted, as there
may be situations in which the minimizing density profile $\{\rho(x)\}$ is
on the boundary of the permissible domain, e.g., one might have
$\rho_A(x)=0$ for some values of $x$, and when that happens the minimizing
profile need not satisfy the ELE.  In the proof of existence we will
consider the weak $L^1$ topology on profiles, in which a sequence $\{f_k\}$
converges to $f$, where $f_k,f\in L^1([0,1])$, if and only if
 \be \label{eq:weak}
   \lim_{k\to\infty}\int_0^1 dx \,f_k(x)\phi(x) = \int_0^1dx\, f(x)\phi(x)
 \ee
 for all bounded and measurable functions $\phi$.  We will use the fact
that, by the Dunford-Pettis Theorem \cite{DP}, the subset 
${\cal K} := \{ f\mid 0 \le f(x) \le 1\ {\rm a.e.}\}$ of $L^1([0,1])$
is compact in this topology (since it is clearly uniformly integrable). We
need also one other preliminary result.

\begin{lem} \label{lem:lsc} The free energy $\F$ is lower semicontinuous in
the weak $L^1$ topology, that is, if $n=(n_{A},n_{B},n_{C})$ and
$n_k=(n_{k,A},n_{k,B},n_{k,C})$, $k=1,2,\ldots$, are profiles with each
sequence $\{n_{k,\alpha}\}$ converging to $n_\alpha$ in the sense of
\eqref{eq:weak} then $\liminf_{k\to\infty}\F(\{n_k\})\ge\F(\{n\})$.\end{lem}

\begin{proof}  It suffices to show that if $\lim_{n\to\infty}f_n = f$ and
$\lim_{n\to\infty}g_n = g$ in the weak $L^1$ topology then
 \be \label{eq:first}
  \lim_{n\to\infty}\int_0^1 dx\, \int_0^1 dz\,\Theta(z-x)f_n(x)g_n(z)  
     =  \int_0^1 dx\, \int_0^1 dz\, \Theta(z-x)f(x)g(z) 
 \ee
and
 \be  \label{eq:second}
 \liminf_{n\to\infty}\int_0^1dx\,  f_n(x)\ln f_n(x)  
    \ge \int_0^1dx\,  f(x)\ln f(x)\ .
 \ee
To verify \eqref{eq:first}, define $h_n(x) = \int_0^1 dz \Theta(z-x)f_n(x)$
and $h(x) = \int_0^1 dz \Theta(z-x)f(x)$, and notice that by the definition
of weak convergence,
 \be \lim_{n\to\infty}h_n(x) = h(x) 
 \ee
 almost everywhere. Then as each $h_n$ takes its values in $[0,1]$, the
dominated convergence theorem implies that $h_n$ converges to $h$ strongly
in the $L^1$ norm. Fix $\epsilon>0$, and pick $N$ so that
$\|h_n - h\|_1 < \epsilon$ for all $n\ge N$. Then
 \bea 
\left| \int_0^1 dx\, g_n(x)h_n(x) -  \int_0^1 dx\, g(x)h(x)\right|&&\cr 
   &&\hskip-180pt \le\left|\int_0^1 dx\, (g_n(x)- g(x))h(x)\right|
    + \left|\int_0^1 dx\, g_n(x)(h_n(x)- h(x))\right|.\quad
  \eea
 But
$\left|\int_0^1
dx\, g_n(x)(h_n(x)- h(x))\right| \le \|g_n\|_\infty\|h_n - h\|_1 \le
\epsilon$ since $0 \le g_n \le 1$. Likewise, since $0 \le h \le
1$, $\lim_{n\to \infty}\int_0^1 dx\, (g_n(x)- g(x))h(x) =
0$ by the definition of weak convergence.  Thus for all $n$ sufficiently
large,
$\left|
\int_0^1 dx\, g_n(x)h_n(x) - \int_0^1 dx\, g(x)h(x)\right| <
2\epsilon$, and \eqref{eq:first} follows.

Equation \eqref{eq:second} is an easy consequence of the convexity of the
function $ -S(f)=\int_0^1dx\, f(x)\ln f(x)$ on ${\cal K}$. Indeed, for any
real number $a$, the set $\K_a=\{ f\in {\cal K} : -S(f) \le a\}$ is convex.
It is also closed in the $L^1$ norm topology, for if $\{f_n\}$ is a
norm convergent sequence in $\K_a$ with limit $f$ then there is a
subsequence converging almost everywhere, and passing to the subsequence
and using the dominated convergence theorem, with the obvious bound
$|f_n| \le 1/e$, yields $-S(f) = \lim_{n\to\infty}-S(f_n) \le a$, so that
$f\in \K_a$  as well.  But by a
theorem of Mazur \cite{M}, a convex set in a Banach space is closed if and
only if it is weakly closed, and so $\K_a$ is
weakly closed. It follows easily from this that if $\{f_n\}$ converges
weakly to $f$, then
 \be
  -S(f) \le \liminf_{n\to\infty}-S(f_n)\ ,
 \ee
 which is \eqref{eq:second}.\end{proof}

\begin{proofof}{\cthm{thm:exist}} Let $\{n_{A,k},n_{B,k},n_{C,k}\}$ be a
minimizing sequence for $\F$, with $n_{k,\alpha}\in\K$,
$\sum_\alpha n_{\alpha,k}=1$, and $\int_0^1n_\alpha(x)\,dx=r_\alpha$, so
that
 \be
  \lim_{k\to\infty} {\cal F}(\{n_k(x)\}) 
      = F(r_A,r_B,r_C)
 \ee
By the weak compactness of $\K$ we can choose a subsequence along which
each $n_{\alpha,k}$ converges to some $\rho_\alpha\in\K$. From
\eqref{eq:weak} it follows immediately that
$\int_0^1 dx\,\rho_\alpha(x) = r_\alpha$, from $\sum_\alpha n_{\alpha,k}=1$
that $\sum_\alpha \rho_\alpha=1$, and from \clem{lem:lsc} that
 \be 
   {\cal F}(\{\rho(x)\}) 
    \le \liminf_{n\to\infty}{\cal F}(\{n_{k}(x)\}) 
     = F(r_A,r_B,r_C)\ .
 \ee
 Thus, ${\cal F}(\{\rho(x)\}) = F(r_A,r_B,r_C)$.  

 We next show that each $\rho_\alpha$ is uniformly bounded below by some
constant $\delta>0$.  Suppose for example that this is not the case for
$\rho_A$, so that for each $\delta > 0$, the set
$D_\delta = \{ x\mid\rho_A(x) < \delta\}$ has strictly positive measure
$|D_\delta|$.  We can then ``fill in the hole'' in $\rho_A$: we transfer
some mass from the larger of $\rho_B$ and $\rho_C$ to $\rho_A$, in a way
that preserves the constraint $\rho_A(x)+\rho_B(x)+\rho_C(x) =1$, and then
transfer mass in the opposite direction on a certain ``safe'' set $M$ to
restore the constraints $\int_0^1\rho_\alpha\,dx=1$.  Because ``entropy
abhors a vacuum'', these transfers will, for $\delta$ sufficiently small,
strictly lower the value of $\F$, thus  contradicting the optimality of
$\{\rho_\alpha\}$.
 
On the set $M$, $\rho_A$ should be strictly bounded below; we take
$M = \{x \mid \rho_A(x) > r_A/2\}$ and note that $|M|>0$ because $\rho_A$ has
mean value $r_A$.  Next, let $E_\delta$ be any subset of $D_\delta$
with $0<|E_\delta| \le |M|$, let $F_\delta$ be the subset of
$E_\delta$ on which $\rho_B \ge \rho_C$, and let
$G_\delta = E_\delta\backslash F_\delta$.  Finally, define
$\{\widetilde \rho_\alpha\}$ by
 \bea
 \widetilde \rho_A(x) &=& \rho_A(x) + \delta\, 1_{E_\delta}(x) 
        - \delta\frac{|E_\delta|}{|M|}\,1_M(x)\ ,\\
 \widetilde \rho_B(x) &=& \rho_B(x) - \delta\, 1_{F_\delta}(x) 
      +  \delta\frac{|F_\delta|}{|M|}\,1_M(x)\ ,\\
  \widetilde \rho_C(x) &=& \rho_C(x) - \delta\, 1_{G_\delta}(x) 
      +  \delta\frac{|G_\delta|}{|M|}\,1_M(x)\  .
 \eea
 Using the fact that, on $D_\delta$,
$\max\{\rho_A(x)\,\ \rho_B(x)\} \ge (1-\delta)/2$, and on $M$,
$\rho_B(x), \rho_C(x) \le 1 - \rho_A/2 < 1$, one sees easily that if
$0<\delta<\min\{r_A/2,1/3\}$ then $\widetilde \rho_\alpha$
satisfies $0\le\widetilde \rho_A(x)\le 1$.

  It is now easy to verify  that for all $\delta$ sufficiently small,
 \be
   -S(\widetilde \rho_A) = -S(\rho_A)
     + \bigl(\delta\ln(\delta)+O(\delta)\bigr)|E_\delta|.
 \ee
  Moreover, changes in $S(\rho_B)$, $S(\rho_C)$, and
the energy components of $\F$ are of order $O(\delta)|E_\delta|$, and thus,
for $\delta$ sufficiently small,
${\cal F}(\{\widetilde \rho_\alpha\}) < {\cal F}(\{ \rho_\alpha\})$, which
contradicts the optimality of $\{ \rho_\alpha\}$. Hence for some
$\delta>0$, it must be the case that $D_\delta$ is a null set.  The same
argument applies of course to $\rho_B$ and $\rho_C$, and hence each of the
$\{\rho_\alpha\}$ is uniformly bounded below by a strictly positive
constant.  Then from the fact that $\rho_A(x)+\rho_B(x)+\rho_C(x) =1$
almost everywhere, it follows that $\rho_\alpha(x) < 1-2\delta$ for almost
every $x$ and each $\alpha$.

Finally, since we have an interior minimum, $\{\rho_\alpha\}$ must
satisfy the ELE for $\F$, that is, $\F_A-\F_C$ and $\F_B-\F_c$ must be
constant (see \eqref{ELE0}), which from \eqref{eq:FA} implies that
$\rho_A/\rho_C$ and $\rho_B/\rho_C$ are differentiable.  From this the
constraint $\sum_{\alpha}\rho_\alpha(x)=1$ implies readily that all the
$\rho_\alpha$ are differentiable, and the argument of
Section~\ref{sec:scaling} them shows that they satisfy the ELE in the
form \eqref{eq:int}.  This implies infinite differentiability.  
\end{proofof}
  
\section{Discussion and Conclusion}\label{sec:conclusion}

The dynamics defining the weakly asymmetric ABC model on the interval are
entirely local and identical to those for the model on a ring, except for
the insertion of a barrier between sites $N$ and 1.  Remarkably, when
$N_A=N_B=N_C=N/3$ this has no effect on the stationary state of the system,
which (as was known before for the ring) is given by a Gibbs measure with
mean field type asymmetric long range pair interactions (see
\eqref{eq:2.2}, \eqref{eq:2.3}).  When the species numbers are unequal,
however, the stationary state on the ring is a nonequilibrium one, with a
net current, while the state on the interval is again given by a 
canonical Gibbs measure with the same interactions.

In the scaling limit $N\to\infty$,
$N_\alpha/N\to r_\alpha$, the system on the ring with $r_A=r_B=r_C=1/3$ was
known to have a second order transition at $\beta=\beta_c=2\pi\sqrt3$ from
a macroscopic state with uniform density profiles $\rho_\alpha(x)=1/3$ to
one in which the density profiles are periodic, with period one.  This
transition, which by the isomorphism between the ring and interval systems
at equal densities carries over to the interval, corresponds to a broken
symmetry in which the phase of the typical density profile in the scaling
limit is uniformly distributed over $[0,1]$.  This corresponds in
\eqref{eq:superpos} to $\Omega=[0,1]$, with the measure $\kappa$ uniform on
$\Omega$. 

In the case of the ring there is strong evidence \cite{BDLvW}, but no proof
so far, that there will be a transition from uniform to nonuniform typical
configurations also for a range of unequal, but strictly positive,
densities.  It might then be natural to conjecture that something similar
happens on the interval.  Our results here, however, prove that this is not
the case: for unequal densities the change from $\beta=0$ to
$\beta=\infty$ is smooth, with unique density profiles $\rho_\alpha(x)$ for
each $\beta<\infty$.  In the limit $\beta\to\infty$ the system will
segregate into three or four blocks of different species; see the
discussion in Section~2.1 of the ground states of the finite system.  When
$0<r_B<r_A=r_C$ the limiting configuration has four blocks of particles,
$\bf BCAB$, with the $B$ particles evenly divided between the left and
right ends of the interval, that is, the degeneracy of the ground state of
the finite system, in which the $B$ particles could be arbitrarily divided
between the two ends, is broken in the scaling limit when $\beta\to\infty$.

A striking characteristic of the resulting $(T,r_A,r_B)$ phase diagram
(Fig.~7) is that one observes a phase transition at precisely one value of
the external parameters (here $r_A$ and $r_B$) as the temperature is
lowered.  Such a feature is typically encountered for grand canonical
ensembles of systems in the presence of an ordering field (e.g.  the
$(T,H_x,H_y)$ phase diagram of the $XY$ model in a magnetic field
$\vec{H}=(H_x,H_y)$).  Note, however, that here we are considering a mean
field model and permitting only those changes in the system in which the
densities $r_A$, $r_B$, and $r_C$ are kept strictly fixed.  The question of
what happens for more general variations, that is, if we consider a grand
canonical ensemble, has to be investigated separately (see note~5 below).
This is unlike the case of systems with short range interactions, where one
has an equivalence of ensembles.  What is interesting in this model
compared with the usual mean field situation is the rich spatial structure
of the equilibrium states, which is due entirely to the directional
asymmetry of the interactions.  Thus if we modified the energy function in
\eqref{eq:2.3} to involve a sum over all $j\ne i$, not just $j>i$, we would
obtain $E_N^{\rm mf}=N_AN_B+N_AN_C+N_BN_C$ which, in the canonical
ensemble, would give uniform minimizing density profiles at all
temperatures, since the interaction would not depend on the spatial
structure and the entropy term prefers the uniform state.

 As we have shown, the minimizing profiles are given by pieces of the
graphs of elliptic functions $y_K$ describing the $\tau$-periodic
trajectory of a particle with zero energy in a quartic confining potential.
It follows from the general analysis of the Euler-Lagrange equations that
these elliptic functions have the property that
$y_K(t)+y_K(t+\tau/3)+y_K(t-\tau/3)=1$ while
$y_K(t)y_K(t+\tau/3)y_K(t-\tau/3)=K$ is independent of $t$.  It is not
clear to us whether or not this property of these elliptic functions
$y_K(t)$ is known in the literature.

We note that, as pointed out in Section~\ref{sec:model}, the local measures
$\mu_x$ must be either product measures or superpositions of product
measures.  Clearly when \eqref{eq:2.6newa} holds, as it will wherever the
minimizers are unique, $\mu_x=\nu_{\rho(x)}$ will be a product measure with
density $\rho(x)$, while for $r_A=r_B=r_C=1/3$ and $\beta>\beta_c$ the
$\mu_x$ will be a superposition of the measures $\mu_\rho(z)$ as $z$ varies
uniformly over $[0,1]$.

We end this section with several open problems.
 
 \smallskip\noindent
 1. It follows from our analysis that for large $\beta$ there will be
solutions of the ELE of type $n>1$ which are stationary points, but not
global minimizers, of $\F$.  We have not determined whether or not these
correspond to local minima.

 \smallskip\noindent
 2. It follows from the proof (see Section~\ref{sec:convex}) of the strict
convexity of $\F\{n(x)\}$ for small $\beta$ that the fluctuations about the
minimizing $\rho(x)$ at fixed $r_\alpha$ are (constrained) Gaussian.  We
know that this is not true for $\beta>\beta_c$ and $r_A=r_B=r_C=1/3$, due
to the existence of a one parameter family of minimizing profiles.  What of
the fluctuations about minimizing densities for other values of $r_\alpha$
when $\beta$ is not very small?

\smallskip\noindent
 3. As mentioned in the introduction, Evans et al.~considered the ABC model
on the ring with general rates $q_{\alpha,\gamma}$ $(\alpha\ne\gamma)$ and
found that the stationary state is Gibbsian whenever
$N_\alpha^{-1}\log(q_{\alpha+1,\alpha+2}/q_{\alpha+2,\alpha+1})$ is
independent of $\alpha$.  This is possible, of course, only when the ratio
of any two of these logarithms is rational, and then only when $N$ is
a multiple of some smallest possible system size, just as for the model
considered in the body of this paper the stationary state can be Gibbsian
only if $N$ is divisible by 3.  To study the weakly asymmetric scaling
limit one would take
$q_{\alpha,\alpha+1}=e^{-\beta c_\alpha/N}q_{\alpha+1,\alpha}$ with some
normalization, say $\sum_\alpha c_\alpha=1$.  The energy
 \be
E_N  = {\frac{1}{N}} \sum^{N-1}_{i=1} \sum^{N}_{j=i+1} \sum_\alpha 
    c_\alpha\eta_{\alpha+1} (i) \eta_\alpha(j). \label{Ec}
 \ee
 would then be translation invariant when $N_\alpha=c_\alpha N$.  As in the
current paper, however, one may use \eqref{Ec} to obtain a Gibbs measure
$Z^{-1}\exp[-\beta E_N]$ on the interval for any values of the $N_\alpha$, and
then study the scaling limit \eqref{hslimit}.  It would be interesting to
carry out an analysis of the free energy minimizing profiles also for this
model.

 \smallskip\noindent
 4. The ABC model can be extended to higher dimensions either on a torus or
in a box \cite{KBEM}. There does not, however, seem to be any natural way
to extend the Gibbs measure to these systems for any values of the
$r_\alpha$'s.

 \smallskip\noindent
 5. Our analysis of $\F(\{n\})$ has been carried out entirely for the
canonical ensemble, the natural one for the exchange dynamics.  From the
point of view of the Gibbs measure \eqref{eq:2.2} it is natural to consider
also the grand canonical ensemble in which the $N_\alpha$ are not fixed.
In that case one would have to consider variations in the grand canonical
partition function, the pressure, for fixed chemical potentials.  It turns
out that in this case there is a phase transition at values of $\beta$ less
than $\beta_c$.  We have also found a natural generalization of the ABC
dynamics which leads to a grand canonical Gibbs measure: one considers the
system on a ring of $N+1$ sites, with one special particle (of type $X$)
replacing the boundaries in the interval model, and permits exchanges
$AX\leftrightarrow XB$, and cyclic permutations of these, at rates which
satisfy detailed balance with the given chemical potentials.  This will be
the subject of future work.

\medskip
\noindent {\bf Acknowledgments:} We thank Lorenzo Bertini, Thierry
Bodineau, Bernard Derrida, Erel Levine and Errico Presutti for helpful
discussions.  The work of J.L.L.~and A.A.~was supported by NSF Grant
DMR-0442066 and AFOSR Grant AF-FA9550-04.  Support of the Israel Science
Foundation (ISF), the Minerva Foundation with funding from the Federal
Ministry for Education and Research and of the Albert Einstein Center for
Theoretical physics is gratefully acknowledged. 

\appendix
\section{The profiles as elliptic functions\label{sec:elliptic}}

The function $t(y)$ defined in \eqref{solution} is an elliptic function,
since $U_K(y)$ is a quartic polynomial in $y$. To identify this function
explicitly we introduce a fractional linear or M\"obius transformation
which sends the four roots $0$, $a$, $b$, and $c$ of $U_K(y)$ to the four
roots of the polynomial which appears in the incomplete elliptic integral
of the first kind \cite{ae},
\be
F(x,k) = \int_0^x \frac{dt}{\sqrt{(1-x^2)(1-k^2x^2)}}.
\ee
 We choose the parameter $k$ to satisfy $k \geq 1$ and map
$(0,a,b,c) \to (-1,-1/k,1/k,1)$. The formula for the transformation is
\be \label{mtr}
y \to z = f(y) = \frac{\alpha_+y -1}{\alpha_-y+1},
\ee
 where
\be
\alpha_\pm = \frac{ \pm a b + \sqrt{a b (c-b) (c-a)}}{a b c},
\ee
which implies
\be
k = \frac{1+\alpha_- a}{1-\alpha_+ a}.
\ee
 Writing the inverse M\"obius transformation of \eqref{mtr} as 
 \be \label{invmob}
z \to y = g(z) = \frac{1+z}{\alpha_+ - \alpha_- z},
 \ee
we can write \eqref{solution} as
 \be
\begin{split}
t &= \int_{-1/k}^{f(y)} \frac{(\alpha_+ + \alpha_-) dz}{(\alpha_+ -
\alpha_- z)^2 \sqrt{-2U_K(g(z))}} \\
&= \varkappa \int_{-1/k}^{f(y)} \frac{dz}{\sqrt{(1-z^2)(1-k^2 z^2)}} \\
&= \varkappa \left[ F \left( \frac1k, k \right) + F(f(y),k) \right],
\end{split}
 \ee
where
 \be
\varkappa = \frac{2(\alpha_+ + \alpha_-)}{\sqrt{(1- \alpha_+ c) (1- \alpha_+
b) (1- \alpha_+ a)}}.
 \ee
The period of oscillation \eqref{period} is then
 \be
 \tau = \frac{4}{\varkappa} F \left( \frac1k, k \right).
 \ee

The inverse of the elliptic integral $F(x,k)$ is the Jacobi
elliptic function $\sn(x,k)$, that is, $F(\sn(x,k),k) = x$. 
Then $f(y(t)) = \sn(\varkappa t - F \left( \frac1k, k
\right),k)$, so that 
 \be
y(t) = \frac{1 + \sn(\varkappa t - F \left( \frac1k, k \right),k)}{\alpha_+ -
  \alpha_- \sn(\varkappa t-F \left( \frac1k, k \right),k)} 
 \ee
is the required solution to \eqref{osc}. The density profiles are then
 \be \label{soln}
\begin{split}
\rA(x) &=  \frac{1 + \sn(2\varkappa \beta x- \frac 73 F \left( \frac1k, k
  \right),k)}
{\alpha_+- \alpha_- \sn(2\varkappa \beta x- \frac 73 F \left( \frac1k, k
  \right),k)}, \\ 
\rB(x) &=  \frac{1 + \sn(2\varkappa\beta x-F \left( \frac1k, k
  \right),k)}{\alpha_+ - 
  \alpha_- \sn(2\varkappa \beta x-F \left( \frac1k, k \right),k)},\\ 
\rC(x) &=  \frac{1 + \sn(2\varkappa \beta x+ \frac 13 F \left( \frac1k, k
  \right),k)}
{\alpha_+  - \alpha_- \sn(2\varkappa \beta x +\frac 13 F \left( \frac1k, k
  \right),k)}.  
\end{split}
 \ee

As a check we have derived expressions for values of the elliptic functions
at the special points $\frac 13 F\left( \frac1k, k \right)$ and
$-\frac{7}{3}F \left( \frac1k, k \right)$ in a way analogous to computing
$\sin 5 \pi/12$ (i.e., by showing that the desired value is the root of a
polynomial obtained using trigonometric addition formulae), and then used
the addition formulae of the elliptic functions \cite{ae} to verify
directly the constancy of the sum \eqref{eq:sum} and the product
\eqref{eq:product} of the densities. These computations were very intensive
and were assisted with the help of computer algebra package,
Maple\texttrademark.

\section{Proofs for \crem{rem:trivial}}\label{proofs}

\begin{proofof}{\crem{rem:trivial}} (b) It is convenient to change
variables in the integrals occurring in \eqref{period} and
\eqref{perints1}--\eqref{perints2} by writing $y=a+(b-a)s$, so that for
example \eqref{period} becomes
 \be\label{period1}
\tau_K=4\int_0^1\frac{ds}{\sqrt{a+s(b-a)}\sqrt{s(1-s)}\sqrt{c-(a+s(b-a))}}.
 \ee
 Then as $K\nearrow1/27$ the points $a$, $b$, and $c$ approach $1/3$,
$1/3$, and $4/3$, respectively, so $\tau_K$ approaches
$4\sqrt3\int_0^1ds/\sqrt{s(1-s)}=4\pi\sqrt3$.  Under this change of
variable the integral in \eqref{perints1} becomes (assuming that $K$
is so small that $a<\epsilon$)
 \be\label{I1}
I_1\equiv2\int_0^{\epsilon(K)}
  \frac{ds}{\sqrt{s(s+a)}}
\sqrt{\frac{s+a}{(a+s(b-a))(1-s)(c-(a+s(b-a)))}}
 \ee
 where $\epsilon(K)=(\epsilon-a)/(b-a)$.  As $K\searrow0$, $a=4K+O(K^2)$
and $b_{\pm}=1\pm2\sqrt K+O(K)$, so that $e(K)=\epsilon+O(K)$ and the
second square root in \eqref{I1} is $\eta(s)=1+O(K)$ uniformly on the
integration range, from which
 \be
\lim_{K\searrow0}\frac{I_1}{\ln(1/K)}
 =\lim_{K\searrow0}\frac2{\ln(1/K)}\int_0^\epsilon\frac{ds}{\sqrt{s(s+a)}}=2.  
 \ee
 The argument for the limit in \eqref{perints2} is similar; the
 difference in the limiting values arises because $1-b$ is of order
 $\sqrt K$ for $K$ small. Finally, a similar analysis shows that 
 \be
\int_\epsilon^{1-\epsilon}\frac{dy}{\sqrt{-2U_K(y)}}
 \ee
 is bounded as $K\searrow0$ which, with \eqref{period1}, shows that 
$\lim_{K\searrow0}(\tau_K/\ln(1/K))=6$.

 \smallskip\noindent
 (c) The proof is long, computational, and unilluminating.  We show that
$d\tau_K/dK>0$; as mentioned in Section~\ref{view}, it is most convenient to
consider $a\in(0,1/3)$ as the fundamental parameter and show that
$d\tau_{K(a)}/da>0$ (recall that $K(a)=a(1-a)^2/4$ so $dK/da>0$).  Our
starting point is \eqref{period1}; writing
 \be
h(a,s)\equiv(a+s(b-a))(c-(a+s(b-a)))
 \ee
  and $R=\sqrt{a(4-3a)}$ we find that
 \be\label{perioddiff}
\frac{d\tau_K}{da}
   =4\int_0^1\frac{d\,h(a,s)}{ds}\frac{ds}{h(a,s)^{3/2}\sqrt{s(1-s)}}
   =\frac4R\int_0^1\frac{g(a,s)\,ds}{h(a,s)^{3/2}\sqrt{s(1-s)}},
 \ee
 where
 \be
g(a,s)=3a(1-a)(1+s)+(9a^2-12a+2)s^2+(1-3a-(5-9a)s+(2-3a)s^2)R.
 \ee

Unfortunately, $g(a,s)$ is not nonnegative for all $a,s$, but the sum of
$g$ and its reflection around the point $s=1/2$ is.  This follows from the
next lemma, whose proof is given immediately below. 
 
\begin{lem}\label{lemma:gpos} For $0\le a\le 1/3$ and $0\le s\le 1/2$,
(i)~$h(a,s)\le h(a,1-s)$; (ii)~$g(a,s)\ge0$, and
(iii)~$g(a,s)+g(a,1-s)\ge0$.
\end{lem}

 \smallskip\noindent
 Using the lemma, we have from \eqref{perioddiff}
 \be\begin{split}\label{bige}
  \frac{d\tau_K}{da}
   &=\frac4R\int_0^{1/2}\left[\frac{g(a,s)}
    {h(a,s)^{3/2}}+\frac{g(a,1-s)}{h(a,1-s)^{3/2}}\right]
    \frac{ds}{\sqrt{s(1-s)}}\\ 
  &\ge \frac4R\int\limits_{\ts\atop{0\le s\le 1/2}    
     {g(a,1-s)\le0}}\left[\frac{g(a,s)}
    {h(a,s)^{3/2}}+\frac{g(a,1-s)}{h(a,1-s)^{3/2}}\right]
    \frac{ds}{\sqrt{s(1-s)}}\\
  &\ge \frac4R\int\limits_{\ts\atop{0\le s\le 1/2}
       {g(a,1-s)\le0}}\bigl[g(a,s)+g(a,1-s)\bigr]
    \frac{ds}{h(a,s)^{3/2}\sqrt{s(1-s)}}\\
  &\ge0,
\end{split}\ee
 where the three inequalities in \eqref{bige} are justified by parts (ii),
 (i), and (iii) of the claim, respectively.  Note also that the first
 inequality is strict unless $g(a,s)\le0$ for all $s\in[1/2,1]$, in which
 case the last inequality will be strict (here we use the fact that neither
 $g(a,s)$ nor $g(a,s)+g(a,1-s)$ vanish identically in $s$ for any $a$);
 thus \eqref{bige} actually implies $d\tau_K/da>0$. 
\end{proofof}

\begin{proofof}{\clem{lemma:gpos}}  For (i), note that
 \be
h(a,1-s)-h(a,s)=(b-a)(1-2s)(R-a);
 \ee
 each of the factors on the right hand side is easily seen to be
nonnegative for the range of variables in question.  Before proceeding to
(ii) and (iii) we make two preliminary observations.  First, the
coefficient 
 \be
\lambda(a)\equiv(9a^2-12a+2)+(2-3a)R
 \ee
  of $s^2$ in $g(a,s)$
(which is a quadratic polynomial in $s$) is positive for all $a\in[0,1/3)$,
for from $\lambda(0)=2$, $\lambda(1/3)=0$, and $\lambda'(1/3)=-8$ we see
that a zero of $\lambda$ in the open interval $(0,1/3)$ would imply two
roots of $\lambda''$ there, contradicting $\lambda'''(a)=48/R^5$ strictly
positive.  Second, $\kappa(a)=g(a,1/2)$ is positive on $[0,1/3)$, for
$\kappa'''(a)=-6/(a^2R(4-3a))$ is negative on $(0,1/3)$ and since, as one
easily checks, $\kappa''(1/2)=\kappa'(1/2)=\kappa(1/2)=0$, we conclude that
$\kappa''(a)$, $-\kappa'(a)$, and $\kappa(a)$ are all positive.  These
observations immediately imply (iii) of the claim, since they imply that
$g(a,s)+g(a,1-s)$ is for any $a$ a parabola in $s$ with positive minimum at
$s=1/2$.

We turn finally to (ii) of the claim.  For fixed $a$ the minimum of
$g(a,s)$ occurs at
 \be
s_0(a)=\frac{3a(1-a)+(5-9a)R}{2(9a^2-12a+2+(2-3a)R)}.
 \ee
 For those $a$ for which $s_0(a)\ge1/2$, the observations in the preceding
paragraph imply $g(a,s)\ge0$, $s\in[0,1/2]$.  But $s_0$ is an increasing
function of $a$, since
 \be
s_0'(a)=\frac{5(2-3a)+18a^2(2-a)+3R}{R(9a^2-12a+2+(2-3a)R)^2};
 \ee
 moreover, if $a_0=2/25$ then $s_0(a_0)=(107\sqrt{47}-69)/(88\sqrt{47}+686)>1/2$. Thus it
suffices to show that
 \be\label{gaa}
g(a,s_0(a))=\frac{a(-22+45a-36a^2+9a^3)+(4-9a+18a^2-9a^3)R}
   {2(9a^2-12a+2+(2-3a)R)}
 \ee
 is nonnegative for $a\in(0,a_0)$.  The denominator in \eqref{gaa} is
easily seen to be positive in this range. We denote the numerator by
$\phi(a)$; then $\phi(0)=0$ and $\phi(a_0)>0$, and positivity will follow
once we prove that there is an $a_1$ with $0<a_1<a_0$ such that
$\phi'(a)>0$ for $a\in (0,a_1$ and $\phi'(a)<0$ for $a\in(a_1,a_0)$. But if
$\psi(a)=R\phi'(a)$ then 
 \be
\psi(a)=108a^4-288a^3+234a^2-66a+8+(36a^3-108a^2+90a-22)R
 \ee
  then, as is easily checked, $\psi(0)>0$, $\psi(a)<0$, and $\psi'(a)<0$ on
$(0,a+0)$, since $\psi'(a)=(\psi_1(a)+\psi_2(a)R)/R$, where 
 \be
\psi_1(a)=-22+303a-810a^2+738a^3-216a^4
 \ee
 and
 \be
\psi_2(a)= -33-432a^2+234a+216a^3,
 \ee
 when written as polynomials in $a-a_0$, have negative constant
  term and alternating signs, and hence are negative on $(0,a_0)$.
\end{proofof}


\begin{thebibliography}{99}

\bibitem {Leib}E. H. Lieb, D. C. Mattis (editors), {\it Mathematical Physics
in One Dimension,} Academic Press, New York (1966).

\bibitem{Percus} J.~K.~Percus, {\em Exactly solvable models of classical
many-body systems},  in  Simple models of equilibrium and nonequilibrium
  phenomena, ed. J.~L.~Lebowitz, Amsterdam, North-Holland, 1987.

\bibitem {Privman} V. Privman (editor), {\it Nonequilibrium statistical
mechanics in one dimension,} Cambridge University Press, Cambridge, 1997.

\bibitem{Schutz} G.~M.~Sch\"utz, {\em Exactly solvable models for
  many-body systems 
  far from equilibrium},  in Phase Transitions and Critical Phenomena,
    {\bf 19}, ed. C.~Domb and J.~L.~Lebowitz, Academic Press, London, 2000.

\bibitem{Evans98} M.R. Evans, Y. Kafri, H.M. Koduvely, and D.  Mukamel,
{\em Phase separation in one-dimensional driven diffusive systems},
\PRL~{\bf 80}, 425--429 (1998); {\em Phase separation and coarsening in
one-dimensional driven diffusive systems: Local dynamics leading to
long-range Hamiltonians}, \PRE~{\bf 58}, 2764--2778 (1998).

\bibitem{LBR} R. Lahiri, M. Barma, and S. Ramaswamy, {\em Strong phase
separation in a model of sedimenting lattices}, Phys. Rev. E {\bf 61},
1648---1658 (2000).

\bibitem{CDE} M. Clincy, B. Derrida, and M.R. Evans, {\em Phase transition in
the $ABC$ model}, \PRE~{\bf 67}, 066115 (2003).

\bibitem{BDLvW} T. Bodineau, B. Derrida, V. Lecomte, and F. van Wijland,
{\em Long range correlations and phase transition in non-equilibrium
diffusive systems}, J.  Stat. Phys. {\bf 133}, 1013--1031 (2008).

\bibitem{BDGJ} L. Bertini, A. De Sole, D. Gabrielli, G. Jona-Lasinio, C.
Landim, {\em Towards a nonequilibrium thermodynamics: a self-contained
macroscopic description of driven diffusive systems}, J.  Stat. Phys {\bf
135}, 857--872 (2009).

\bibitem{FF} G. Fayolle and C. Furtlehner, {\em Stochastic deformations of
sample paths of random walks and exclusion models}, in Mathematics and
Computer Science III: Algorithms, Trees, Combinatorics and Probabilities
(Trends in Mathematics) , ed. M. Drmota, P. Flajolet, D. Gardy, and B.
Gittenberger, Birkh\"auser, Basel, 2004; G. Fayolle and C. Furtlehner, {\em
Stochastic dynamics of discrete curves and multi-type exclusion processes},
J. Stat. Phys. {\bf 127}, 1049--1094 (2007).

\bibitem{SS}  S. Sandow and G. Sch\"utz, {\em On $U_q[SU(2)]$-symmetric
  driven diffusion},  Europhys. Lett {\bf 26}, 7--12 (1994).

\bibitem{BE} R. A. Blythe and M. R. Evans, {\em Nonequilibrium steady states of
  matrix-product form: a solver's guide},  J. Phys. A {\bf 40}
R333  (2007).


\bibitem{AL} M. Aizenmann and E. Lieb, {\em The III$\,{}^{\rm rd}$ Law of
  Thermodynamics and the Degeneracy of Ground States in Lattice
  Systems}, J. Stat. Phys. {\bf 24}, 279--297 (1981).

\bibitem{Ligg} T.~M.~Liggett, {\em Stochastic interacting systems: contact,
voter and exclusion processes}, Springer-Verlag, Berlin, 1999.

\bibitem{HS} E. Hewitt and L.~J.~Savage, {\em Symmetric measures on Cartesian
products}, Trans.~Amer.~Math.~Soc. {\bf 80}, 470--501 (1955).

\bibitem{WC} C. Kipnis and C. Landim, {\em Scaling limits of interacting
  particle systems},  Grundlehren der Mathematischen
  Wissenschaften {\bf 320}, Springer-Verlag, Berlin, 1999.

\bibitem{Ising} J.L. Lebowitz, {\em Coexistence of Phases in Ising
  Ferromagnets},  J. Stat. Phys. {\bf 16}, 463--476 (1977).

\bibitem{LD} R.S. Ellis, {\em Entropy, large deviations, and statistical
  mechanics},  Grundlehren  der Mathematischen 
  Wissenschaften {\bf 271}, Springer-Verlag, Berlin, 1985.

\bibitem{DP} N. Dunford and B. J. Pettis, {\em Linear operators on summable
functions}, Trans. Amer. Math. Soc. {\bf 47}, 323--392 (1940).

\bibitem{M} S. Mazur, {\em \"Uber konvexe Menge in linearen normierten Raumen},
Studia Math. {\bf 4}, 70--84 (1933).

\bibitem{KBEM} Y. Kafri, D. Biron, M. R. Evans, and D. Mukamel, {\em Slow
coarsening in a class of driven systems}, Euro. Phys. J. B {\bf 16}, 669--676
(2000).

\bibitem{ae} J.V. Armitage and W.F. Eberlein, {\it Elliptic functions},
  London Mathematical Society Student Texts {\bf 67},
  Cambridge University  Press, Cambridge, 2006. 

\end{thebibliography}
\end{document}